\documentclass[11pt]{article}

\usepackage{microtype}
\usepackage{graphicx}
\usepackage{subcaption}
\usepackage{booktabs} 
\usepackage[top=1in,bottom=1in,left=1in,right=1in]{geometry}
\usepackage{natbib}
\usepackage{algorithm}
\usepackage{algorithmic}

\usepackage{hyperref}
\makeatletter
\let\Hy@linktoc\Hy@linktoc@none
\makeatother

\newcommand{\lp}{\left(}
\newcommand{\rp}{\right)}
\newcommand{\lb}{\left[}
\newcommand{\rb}{\right]}
\newcommand{\lbp}{\left\{}
\newcommand{\rbp}{\right\}}
\newcommand{\lba}{\left\lvert}
\newcommand{\rba}{\right\rvert}
\newcommand{\lV}{\left\lVert}
\newcommand{\rV}{\right\rVert}
\newcommand{\mV}{\middle\Vert}
\newcommand{\mv}{\middle\vert}

\newcommand{\mcal}{\mathcal}

\newcommand{\mbb}{\mathbb}
\newcommand{\msf}{\mathsf}
\newcommand{\la}{\leftarrow}
\newcommand{\ra}{\rightarrow}

\newcommand{\lan}{\langle}
\newcommand{\ran}{\rangle}

\newcommand{\eqDef}{\triangleq}
\newcommand{\diid}{\overset{\text{i.i.d.}}{\sim}}

\newcommand{\E}{\mathbb{E}}
\newcommand{\Var}{\mathsf{Var}}

\renewcommand{\Pr}{\mathbb{P}}

\usepackage[shortlabels]{enumitem}

\usepackage{amsmath}
\usepackage{amssymb}
\usepackage{mathtools}
\usepackage{amsthm}

\usepackage[capitalize,noabbrev]{cleveref}

\theoremstyle{plain}
\newtheorem{theorem}{Theorem}[section]

\newtheorem{lemma}[theorem]{Lemma}
\newtheorem{corollary}[theorem]{Corollary}
\theoremstyle{definition}
\newtheorem{definition}[theorem]{Definition}

\theoremstyle{remark}
\newtheorem{remark}[theorem]{Remark}

\usepackage[textsize=tiny]{todonotes}

\title{\textbf{The Poisson binomial mechanism for secure and private federated learning}   } 
\author{ Wei-Ning Chen$^{\dag}$ \thanks{The early stage of this work is done while on internship at Google.} \and Ayfer \"Ozg\"ur$^{\dag}$\and Peter Kairouz$^{\ddag}$ } 
\date{%
Stanford University$^{ \dag}$, Google Research$^{ \ddag}$\\[2ex]%
\texttt{\{wnchen, aozgur\}@stanford.edu, kairouz@google.com}
}

\begin{document}
\maketitle
\begin{abstract}
We introduce the Poisson Binomial mechanism (PBM), a discrete differential privacy mechanism for distributed mean estimation (DME) with applications to federated learning and analytics. We provide a 
tight analysis of its privacy guarantees, showing that it achieves the same privacy-accuracy trade-offs as the continuous Gaussian mechanism. Our analysis is based on a novel bound on the R\'enyi divergence of two Poisson binomial distributions that may be of independent interest. 

Unlike previous discrete DP schemes based on additive noise, our mechanism encodes local information into a parameter of the binomial distribution, and hence the output distribution is discrete with bounded support. Moreover, the support does not increase as the privacy budget $\varepsilon \ra 0$ as in the case of additive schemes which require the addition of more noise to achieve higher privacy; on the contrary, the support becomes smaller as $\varepsilon \ra 0$. The bounded support enables us to combine our mechanism with secure aggregation (SecAgg), a multi-party cryptographic protocol,  without the need of performing modular clipping which results in an unbiased estimator of the sum of the local vectors. This in turn allows us to  apply it in the private FL setting and provide an upper bound on the convergence rate of the  SGD algorithm. Moreover, since the support of the output distribution becomes smaller as $\varepsilon \ra 0$, the communication cost of our scheme decreases with the privacy constraint $\varepsilon$, outperforming all previous distributed DP schemes based on additive noise in the high privacy or low communication regimes. 
\end{abstract}

\section{Introduction}\label{sec:introduction}
The standard technique for ensuring differential privacy (DP) \cite{dwork2006calibrating} of learning algorithms is to add noise either to the output of a function evaluated on the data (in the centralized setting) or locally to the data itself (in federated settings \cite{kairouz2019advances, mcmahan2016communication}). Two commonly used distributions for noise are the Gaussian and Laplace distributions. While simple enough for mathematical reasoning and analysis, the continuous nature of these distributions presents a number of challenges. First, it is not possible to represent real samples on finite computers, making these mechanisms prone to numerical errors that can break privacy guarantees \cite{mironov2012significance}. Second, they cannot be used in the federated setting where it may be desirable to first locally perturb the data (e.g. the local model update computed by stochastic gradient descent(SGD) iterations) and then use cryptographic primitives such as secure aggregation (SecAgg)  \cite{secagg}  to allow the server to obtain a summary of the local data (such as the mean of local model updates) without having access to individual  information. This combination of local DP and secure aggregation is desirable as it does not rely on the clients' full trust in the server, while potentially achieving the same utility-privacy trade-off as in the centralized case. However, secure aggregation is based on modular arithmetic which is not compatible with the real output from a privatization mechanism that relies on perturbing data with continuous noise.
This has led to an increasing recent interest in mechanisms that perturb the data (or a function of it) with the addition of discrete noise, such as the binomial in \cite{dwork2006our,agarwal2018cpsgd}, the discrete Gaussian in \cite{canonne2020discrete,kairouz2021distributed}, and Skellam noise in \cite{agarwal2021skellam}. 

These additive discrete noise mechanisms however have a few of their own shortcomings. First, when the data itself is continuous,  as in the case of local model updates obtained from SGD iterations in federated learning, it has to be discretized before the addition of discrete noise. This adds quantization noise and complicates analysis.   Second, these distributions have discrete yet unbounded support which means that the privatized data has to go through modular clipping when combined with secure aggregation protocols which operate on a finite group. This is the approach in \cite{kairouz2021distributed} and \cite{agarwal2021skellam}, which focus on developing differentially private federated learning algorithms by using these additive noise mechanisms locally at the clients and then feeding the privatized local updates to the secure aggregation protocol. Upon modular clipping, however, the discrete additive noise becomes no longer zero-mean, and hence the resulting estimator (of the mean of the local model updates) is \emph{biased}. The bias makes it difficult to provide tight convergence guarantees for stochastic first-order optimization methods which rely on this estimate. In contrast, it is usually not difficult to provide a tight convergence analysis when the optimization method has access to a noisy but unbiased estimate of the true mean of the updates. 
Finally, all additive noise mechanism (continuous or discrete) share the common principle of adding more noise to achieve higher privacy, i.e. they require a higher noise variance when higher privacy is desired. This, however, has a direct impact on the communication cost in federated settings when the noise is added locally. Indeed, for all of the above mentioned schemes \cite{kairouz2021distributed, agarwal2021skellam, agarwal2018cpsgd} the communication cost grows \emph{inversely} with the privacy budget; when high privacy is desired nodes need a large bit budget to communicate the large noise they add to their updates. This contradicts the conclusion of \cite{chen2020breaking} which shows that in a federated learning setting without secure aggregation the optimal communication cost can be made to \emph{decrease} with the privacy budget; intuitively we can use less bits in the high privacy regime because we are required to communicate less information about the local data.

In this paper, we develop a novel differential privacy mechanism that does not rely on additive noise. This mechanism, which we call the multi-dimensional Poisson Binomial mechanism (PBM), takes a continuous input, encodes it into the parameter $p$ of a binomial distribution $\msf{Binom}\lp m, p \rp$, and generates a sample from this distribution\footnote{We note that the binomial mechanism proposed in \cite{agarwal2018cpsgd} is an additive noise mechanism (it adds Binomial noise to the data), and while it has a finite output range, it does not provide any  R\'enyi DP guarantees (which is the main focus of this paper as RDP allows for tightly accounting privacy loss across multiple rounds).}. This results in a finite and discrete output in $ \mbb{Z}_m$ which can be easily combined with the integer modular arithmetic in SecAgg, without the need for quantization or modular clipping.
As a result, the estimate (of the average model updates) obtained at the output of SecAgg is \emph{unbiased} leading, to our knowledge, to the first unbiased privacy scheme compatible with SecAgg. Moreover, the communication cost of PBM decreases when the privacy budget $\varepsilon$ decreases. This is because the first parameter $m$ of the distribution $\msf{Binom}\lp m, p \rp$ is linear in the privacy budget $\varepsilon$ and hence the logarithm of it, which dictates the communication budget, decreases to $1$ as $\varepsilon\rightarrow 0$.

\paragraph{Our contributions.}
Our main technical contributions are summarized as follows.
\begin{itemize}
    \item We introduce the multi-dimensional Poisson binomial mechanism, an unbiased and bounded discrete DP mechanism for distributed mean estimation (DME). We provide a tight analysis for its R\'enyi DP (RDP) guarantees showing that it provides the same utility-privacy trade-off as the continuous Gaussian mechanism. As a by-product, our analysis yields a novel bound on the  R\'enyi divergence of two Poisson binomial distributions that can be useful in other applications.

    \item We show that the communication cost of our scheme (defined as the number of bits needed to achieve the accuracy of the centralized DP model) decreases with the privacy budget, as opposed to previous discrete DP schemes \cite{agarwal2018cpsgd, kairouz2021distributed, agarwal2021skellam}. Thus in the high-privacy regime, our scheme uses significantly less bandwidth while still achieving the right order of accuracy.

    \item We combine PBM with distributed SGD and SecAgg in a FL setting and analyze its convergence rate.
\end{itemize}

\subsection{Problem setup and prerequisites}\label{sec:formulation}
In this section, we present the distributed mean estimation (DME) \cite{an2016distributed} problem under differential privacy and SecAgg. Note that DME is closely related to federated learning with SGD, where in each iteration, the server updates the global model by a noisy mean of the local model updates. This noisy estimate is typically obtained by using a DME scheme, and thus one can easily build a distributed DP-SGD scheme (and hence a private FL scheme) from a differentially private DME scheme.

Consider $n$ clients each with local data $x_i \in \mathbb{R}^d$ that satisfies $\lV x_i \rV_2 \leq c$ (one can think of $x_i$ as a clipped local gradient).  A server wants to learn an estimate $\hat{\mu}$ of the mean $\mu \triangleq \frac{1}{n}\sum_i x_i$ after communicating with the $n$ clients.

\paragraph{Secure aggregation.} In order to fully leverage the distributed nature of FL to enhance clients' privacy, the honest-but-curious server collects local data through a \emph{ secure aggregation protocol}. More precisely, each client encodes $x_i$ into a finite additive group $\mcal{Z}$ by computing $Z_i \eqDef \mcal{A}_{\msf{enc}}(x_i)$. The $n$ clients and the server then participate in the SecAgg protocol, so that only $\sum_i Z_i$ can be revealed to the server. Finally, the server computes $\hat{\mu}$ based on $\sum_i Z_i$, an estimate  of the true mean. The goal is to jointly design an encoder $\mcal{A}_\msf{enc}$ and an estimator $\hat{\mu}$, such that
\begin{enumerate}
    \item $\hat{\mu}$ satisfies a differential privacy constraint (see Definition~\ref{def:DP} and Definition~\ref{def:RDP} for formal statements).
    \item The per-client communication cost $b=\log \lba \mcal{Z} \rba$ is small.
    \item $\hat{\mu}$ is unbiased (i.e. $\E[\hat{\mu}] = \mu$) 
    and has small mean squared error (MSE): $\E\lb \lV \hat{\mu} - \mu \rV^2_2 \rb$.
\end{enumerate}

Without loss of generality, we will set $\mcal{Z}$ to be $\lp\mbb{Z}_M\rp^l$ for some $l,M \in \mbb{N}$ (where $\mbb{Z}_M$ is the group of integers modulo $M$ equipped with modulo $M$ addition), so $b = l\log M$ is the total number of communicated bits. The summation is coordinate-wise modulo $M$ addition, i.e. 
$$ \msf{SecAgg}\lp \mcal{A}_{\msf{enc}}(x_1),...,\mcal{A}_{\msf{enc}}(x_n) \rp = \sum_i \mcal{A}_{\msf{enc}}(x_i) \msf{ mod} M.$$

\paragraph{Differential Privacy.}
Finally, we introduce the notion of differential privacy \cite{dwork2006calibrating} and  R\'enyi differential privacy (RDP) \cite{mironov2017renyi}. We are mostly interested in developing mechanisms that satisfy RDP, as it allows for tight privacy accounting across training iterations.
\begin{definition}[(Approximate) Differential Privacy]\label{def:DP}
For $\varepsilon, \delta \geq 0$, a randomized mechanism $M$ satisfies $(\varepsilon, \delta)$-DP if for all neighboring datasets $D, D'$ and all $\mcal{S}$ in the range of $M$, we have that 
$$ \Pr\lp M(D) \in \mcal{S} \rp \leq e^\varepsilon \Pr\lp M(D') \in \mcal{S} \rp+\delta,  $$
where $D$ and $D'$ are neighboring pairs if they can be obtained from each other by adding or removing all the records that belong to a particular user.
\end{definition}


\begin{definition}[ R\'enyi Differential Privacy (RDP)]\label{def:RDP}
A randomized mechanism $M$ satisfies $(\alpha, \varepsilon)$-RDP if for any two neighboring datasets $D, D'$, we have that $D_\alpha\lp P_{M(D)}, P_{M(D')} \rp\leq \varepsilon$ where $D_{\alpha}\lp P, Q\rp$ is the  R\'enyi divergence between $P$ and $Q$ and is given by 
$$ D_\alpha\lp P, Q \rp \eqDef \frac{1}{\alpha}\log\lp \E_{Q}\lb \lp \frac{P(X)}{Q(X)} \rp^\alpha \rb \rp.$$
\end{definition}
Note that one can cast RDP to (approximate) DP. See Section~\ref{sec:conversion} for details.

\subsection{Related works}

The closest works to ours are the distributed discrete DP mechanisms cpSGD \cite{agarwal2018cpsgd}, DDG \cite{kairouz2021distributed}, and Skellam \cite{agarwal2021skellam}. Unlike our proposed scheme, these mechanisms achieve differential privacy (DP) \cite{dwork2006calibrating} by adding discrete noise that (1) has a distribution that asymptotically converges to a normal distribution, and (2) are (nearly) ``closed'' under addition. However, since the noise is asymptotically normal, in the high-privacy regimes where $\varepsilon$ is small, the variance of the noise (and hence the communication cost) explodes. In addition, since the noise has infinite range (except for cpSGD\footnote{However, we note that cpSGD only satisfies approximate DP but not Renyi DP, so we can only use strong composition theorems \cite{dwork2010boosting, kairouz16} to account privacy loss.}), one has to perform modular clipping in order to perform SecAgg. This leads to bias that can cause issues for the downstream tasks such as SGD. 

{Another closely related work is \citet{cheu2019distributed}, in which a distributed DP scheme for scalar mean estimation under the \emph{ shuffled} model of DP is proposed. The scheme introduced in \citet{cheu2019distributed} can be viewed as a concatenation of stochastic rounding and randomized response, and although introduced under the context of shuffling instead of secure aggregation (which is considered in this work), statistically their scheme is equivalent to (the scalar version of) our scheme when the parameters are properly selected. 

However, we remark on several main differences as follows. First, under the shuffled model, \citet{cheu2019distributed} suffers from sub-optimal communication cost; indeed, the per-client message size (and hence the communication cost) grows as $O(\sqrt{n}\varepsilon)$, which is improved in follow-up work \cite{ghazi2021power}. This sub-optimal communication issue, however, does not appear under the secure aggregation model, and indeed, the per-client communication cost of our scheme (which is under the secure aggregation model) is $O\lp \log n \rp$. Second, \citet{cheu2019distributed} only considers the scalar mean estimation problem, which does not directly apply to FL. Extending the results to the multi-dimensional setting, while still preserving the order-optimal privacy-accuracy trade-offs, is non-trivial. In this work, we address it via Kashin's representation \citep{lyubarskii2010uncertainty}. Finally, \citet{cheu2019distributed} studies $(\varepsilon, \delta)$ approximate DP and only obtains an order-wise optimal bound on the privacy guarantees. In this paper, we analyze R\'enyi DP (which allows for tighter privacy accounting), and more importantly, our analysis provides a method to numerically compute the \emph{exact} privacy loss. Indeed, in addition to theoretically proving the order-wise optimality (under R\'enyi DP), we numerically show that our scheme converges to the centralized Gaussian mechanism (i.e. it matches the Gaussian mechanism's constants). }

In this paper, we combine our discrete DP mechanism with SecAgg (more precisely, single-server SecAgg) to achieve distributed DP without introducing bias. Single-server SecAgg is achieved via additive masking over a finite group \cite{bonawitz2016practical, bell2020secure}. To achieve provable privacy guarantees, however, SecAgg is insufficient as the sum of local model updates may still leak sensitive information \cite{melis2019exploiting, song2019auditing, carlini2019secret, shokri2017membership}. To address this issue, DP-SGD or DP-FedAvg can be employed \cite{song2013stochastic, bassily2014private, geyer2017differentially, mcmahan2017learning}. In this work, we aim to provide privacy guarantees in the form of  R\'enyi DP \cite{mironov2017renyi} because it allows for tracking the end-to-end privacy loss tightly.

We also distinguish our distributed DP setup from the local DP setup \cite{kasiviswanathan2011can, evfimievski2004privacy, warner1965randomized}, where the data is perturbed on the client-side before it is collected by the server in the clear. Although both local DP and distributed DP with SecAgg do not rely on a fully trusted centralized server, the local DP model provides stronger privacy guarantees as it allows the server to observe individual (privatized) information, while distributed DP requires that the server executes the SecAgg protocol faithfully. Given its strong privacy guarantees, local DP naturally suffers from poor privacy-utility trade-offs \cite{kasiviswanathan2011can,duchi2013local, kairouz16}. That is why we focus on distributed DP via SecAgg in this paper. 

Our scheme also makes use of Kashin's representation \cite{kashin1977section, lyubarskii2010uncertainty}, a powerful tool that enables us to transform the $\ell_2$ geometry of the input data to an $\ell_\infty$ one in a lossless fashion. This facilitates the analysis and allows for decoupling the high-dimensional problem into 1-dimensional sub-tasks. Similar idea has been used in different settings; for instance, \cite{feldman2017statistical, caldas2018expanding, chen2020breaking}.

\section{Main Results}\label{sec:results}
\begin{algorithm}[b]
   \caption{The Poisson Binomial Mechanism}
   \label{alg:pbm}
\begin{algorithmic}
   \STATE {\bfseries Input:} $x_1,...,x_n \in \mcal{B}_d(c)$, parameters $\theta \in [0, \frac{1}{4}]$, $m \in \mbb{N}$, a tight frame $U$ associated with Kashin's representation at level $K>0$
   \FOR{each client $i$}
    \STATE Set $y_i$ to be the Kashin's representation of $x_i$, 
    so $y_i \in \mbb{R}^{\Theta(d)}$ and $\lV y_i\rV_\infty \leq \frac{cK}{\sqrt{d}}$.
    \FOR{coordinate $j$ of $y_i$}
    \STATE $Z_{ij} \la \msf{scalar\_PBM}\lp y_{ij}, m, \theta, c'=\frac{cK}{\sqrt{d}}\rp$
    \ENDFOR
    \STATE Send $Z_i$ to the server via SecAgg
   \ENDFOR
   
    \STATE (Server) Computes $\hat{\mu_y} = \frac{c'}{mn\theta}\lp \sum_i Z_i - \frac{mn}{2} \rp$
    \STATE (Server) Computes $\hat{\mu} = U\hat{\mu_y}$
    \STATE {\bfseries Return:} $\hat{\mu}$
\end{algorithmic}
\end{algorithm}

We introduce the Poisson Binomial mechanism for DME with SecAgg and differential privacy. The proposed protocol (Algorithm~\ref{alg:pbm}) consists of three stages:
\begin{itemize}
    \item Each client computes the Kashin's representation of local data $x_i$ (denoted as $y_i$), which allows for optimally transforming the $\ell_2$ geometry of the data into  $\ell_\infty$.
    \item The $n$ clients apply the scalar Poisson Binomial mechanism (Algorithm~\ref{alg:scalar_pbm}) separately on each coordinate of $y_i$, and the server estimates $\mu_y \eqDef \frac{1}{n}\sum_{i} y_i$.
    \item The server reconstructs $\hat{\mu}_x$ from the Kashin representation of $\hat{\mu}_y$.
\end{itemize}

\begin{algorithm}[tb]
   \caption{The (Scalar) Poisson Binomial Mechanism}
   \label{alg:scalar_pbm}
\begin{algorithmic}
   \STATE {\bfseries Input:} $c > 0$, $x_i \in [-c, c]$, $\theta \in [0, \frac{1}{4}]$, $m \in \mbb{N}$
   \STATE Re-scaling $x_i$:
    $ p_i \eqDef \frac{\theta}{c}x_i+\frac{1}{2}$.
    \STATE Privatization:
    $ Z_i \eqDef \msf{Binom}\lp m, p_i \rp \in \mbb{Z}_m.$
    \STATE {\bfseries Return:} $Z_i$
\end{algorithmic}
\end{algorithm}

Note that Algorithm~\ref{alg:pbm}  builds on Algorithm~\ref{alg:scalar_pbm}, the scalar version of PBM, which we analyze in Section~\ref{sec:scalar_pbm}. Parameters $(m, \theta)$ determine privacy, communication cost, and MSE; in other words, the privacy-utility trade-offs of Algorithm~\ref{alg:pbm} can be fully characterized by $(m, \theta)$, which we summarize in the following theorem.

\begin{theorem}\label{thm:main_informal}
Let $\lV x_i\rV \leq c$, $m \in \mbb{N}$, and $\theta \in [0, \frac{1}{4}]$. Then with parameters $m, \theta$, Algorithm~\ref{alg:pbm}:

\begin{itemize}
    \item satisfies $\lp \alpha, \varepsilon(\alpha) \rp$-RDP for any $\alpha>1$ and $\varepsilon(\alpha) = \Omega\lp{dm\theta^2\alpha}/{n}\rp$,
    \item requires $O(d\lp\log m+\log n\rp)$ bits of per-client communication,
    \item yields an unbiased estimator $\hat{\mu}$ with $O\lp \frac{c^2}{nm\theta^2} \rp$ MSE.
\end{itemize}
\end{theorem}
\begin{remark}
Although in Theorem~\ref{thm:main_informal} we present an asymptotic result, we remark that (1) the MSE can be upper bound explicitly, and (2) the R\'enyi DP can be computed numerically (as shown in Section~\ref{sec:numerical}). Indeed, we show that when we pick $\theta$ small enough, the MSE of PBM converges to the (centralized continuous) Gaussian mechanism quickly.
\end{remark}

Several observations are given in order. First, the privacy guarantee $\varepsilon(\alpha)$ can be written as a function of the variance (i.e., the MSE): $\varepsilon(\alpha) = \Omega\lp \frac{dc^2 \alpha}{n^2\msf{MSE}\lp \hat{\mu} \rp} \rp$. This privacy-accuracy trade-off matches that of the (centralized) Gaussian mechanism \footnote{We notice that when the $\ell_2$ sensitivity is $c^2/n^2$, a Gaussian mechanism that adds $N(0, \sigma^2\mbb{I}_d)$ noise achieves RDP $\varepsilon_{\msf{Gauss}}(\alpha) = \frac{c^2\alpha}{2n^2\sigma^2}$, and the corresponding $\msf{MSE}(\hat{\mu}_{\msf{Gauss}}) = d\sigma^2$.
} given by 
$\varepsilon_{\msf{Gauss}}(\alpha) = \Omega\lp \frac{dc^2\alpha}{n^2\msf{MSE}\lp \hat{\mu}_{\msf{Gauss}} \rp} \rp$ (which is obtained by bounding the sensitivity of the mean function by $c^2/n^2$). This implies that Algorithm~\ref{alg:pbm} attains order-optimal errors. In Section~\ref{sec:numerical} below, we numerically compute the MSE-privacy trade-offs of PBM and the Gaussian mechanism.

 Note that in Theorem~\ref{thm:main_informal}, both $\varepsilon(\alpha)$ and the the variance of the estimator depend on the parameters $m$ and $\theta$ of the algorithm through the product $m\theta^2$. Hence, this leaves some freedom in the choice of $m$ and $\theta$ if one is concerned only with privacy and MSE. However, the choice of $m$ also dictates the communication cost. We next describe how one can pick $(\theta,m)$ to minimize the communication cost for the same $\lp \alpha, \varepsilon(\alpha) \rp$-RDP constraint and MSE, where the latter is dictated by the first according to the above trade-off. Observe that the privacy budget $\varepsilon(\alpha)$ fixes the value of the product $m\theta^2$, so to minimize $m$, and hence the communication cost, we would like to pick $\theta$ as large as possible. However, $\theta$ is restricted to $[0, \frac{1}{4}]$. Therefore we can determine $m$ and $\theta$ by the following two steps:
\begin{enumerate}
    \item Set $m = 1$ and compute the corresponding $\theta$ such that the resulting privacy is $\varepsilon(\alpha)$. If $\theta > 1/4$, clip $\theta$ to 1/4. This leads to $\theta =  O\lp\min\lp \frac{1}{4},  \sqrt{\frac{n\varepsilon(\alpha)}{d\alpha}} \rp\rp$.
    \item Then, we adjust $m$ again according to $\theta$. If $\theta = 1/4$ (i.e. when $\theta$ clipped in the previous step), we set $m = O\lp\frac{n\varepsilon(\alpha)}{d\alpha}\rp$. Otherwise $m=1$. Hence $m$ is upper bounded by $\max\lp 1, O\lp\frac{n\varepsilon(\alpha)}{d\alpha}\rp \rp.$
\end{enumerate}

Plugging the above upper bound on $m$ to Theorem~\ref{thm:main_informal}, the communication cost becomes $
O\lp d\lp \log \lp n +
 \frac{n\varepsilon(\alpha)}{d\alpha} \rp \rp \rp$.

 Next, to compare the communication cost of our scheme with previous schemes, we convert it into $(\varepsilon_\msf{DP}(\delta), \delta)$-DP via Lemma~\ref{lemma:rdp_to_approx_dp_2} 
 and arrive at the following corollary:
 \begin{corollary}[Approximate DP of PBM]\label{cor:pbm_approx_dp}
 By setting $\theta = O\lp \min\lp \frac{1}{4}, \sqrt \frac{n\varepsilon_{\msf{DP}}^2}{d\log(1/\delta)} \rp\rp$ and $m = \lceil\frac{n\varepsilon^2_\msf{DP}}{d\log(1/\delta)} \rceil$, Algorithm~\ref{alg:pbm} satisfies $\lp \varepsilon_\msf{DP}, \delta \rp$-approximate DP. Moreover, the (per-client) communication cost is $O\lp d \lp \log \lp n +\frac{n\varepsilon_{\msf{DP}}^2}{d\log(1/\delta)}\rp \rp\rp$, and $\hat{\mu}$ is unbiased with MSE at most $O_\delta\lp \frac{c^2d}{n^2\varepsilon_{\msf{DP}}^2} \rp$.
\end{corollary}

We remark that the communication cost of PBM decreases as  $\varepsilon_\msf{DP}$) decreases, exhibiting the correct dependency on $\varepsilon_\msf{DP}$ in the high-privacy regime. The communication cost of other discrete DP mechanisms based on additive noise, such as \cite{agarwal2018cpsgd, kairouz2021distributed, agarwal2021skellam} increase as $\varepsilon$ gets smaller. For instance, the DDG mechanism \cite{kairouz2021distributed} requires $O\lp d \lp \log \lp n +\frac{d}{\varepsilon_{\msf{DP}}^2}\rp \rp\rp$ bits of communication per-client, which becomes unbounded when $\varepsilon_\msf{DP} \ra 0$. See Table~\ref{tbl:approx_dp} for a comparison.

\begin{table}[h!]
\small
    \centering
    \begin{tabular}{|c|c | c | c| } 
    \hline
    & communication & MSE & bias   \\ [0.5ex] 
    \hline
    PBM 
    & $O\lp d  \log \lp \lceil {\color{blue}\frac{n\varepsilon_{\msf{DP}}^2}{d}}\rceil\cdot n \rp\rp$ & $O_{\delta}\lp \frac{c^2d}{n^2\varepsilon_{\msf{DP}}^2} \rp$ & {\color{blue}no} \\ 
    \hline
    DDG 
    & $O\lp d \log \lp \lceil\frac{d}{\varepsilon_{\msf{DP}}^2}\rceil\cdot n \rp\rp$ & $O_\delta\lp \frac{c^2d}{n^2\varepsilon_{\msf{DP}}^2} \rp$ & yes \\ 
    \hline
    Skellam 
    & $O\lp d \log \lp\lceil \frac{d}{\varepsilon_{\msf{DP}}^2}\rceil \cdot n\rp\rp$ & $O_\delta\lp \frac{c^2d}{n^2\varepsilon_{\msf{DP}}^2} \rp$ & yes \\ 
    \hline
    Binomial 
    & $O\lp d \log \lp \lceil\frac{d}{\varepsilon_{\msf{DP}}^2}\rceil \cdot n  \rp\rp$ & $O_\delta\lp \frac{c^2d\log(d)}{n^2\varepsilon_{\msf{DP}}^2} \rp$ & yes \\ 
    \hline
    \end{tabular}
    \caption{A comparison of the communication costs and MSEs of different discrete DP schemes. For the communication cost, we hide the dependency on $\log n$ since we are interested in high-dimensional regimes where $d \gg n$.}\label{tbl:approx_dp}
\end{table}

\subsection{Numerical evaluation}\label{sec:numerical}

\begin{figure}[h]
  	\centering
  	\includegraphics[width=0.65\linewidth]{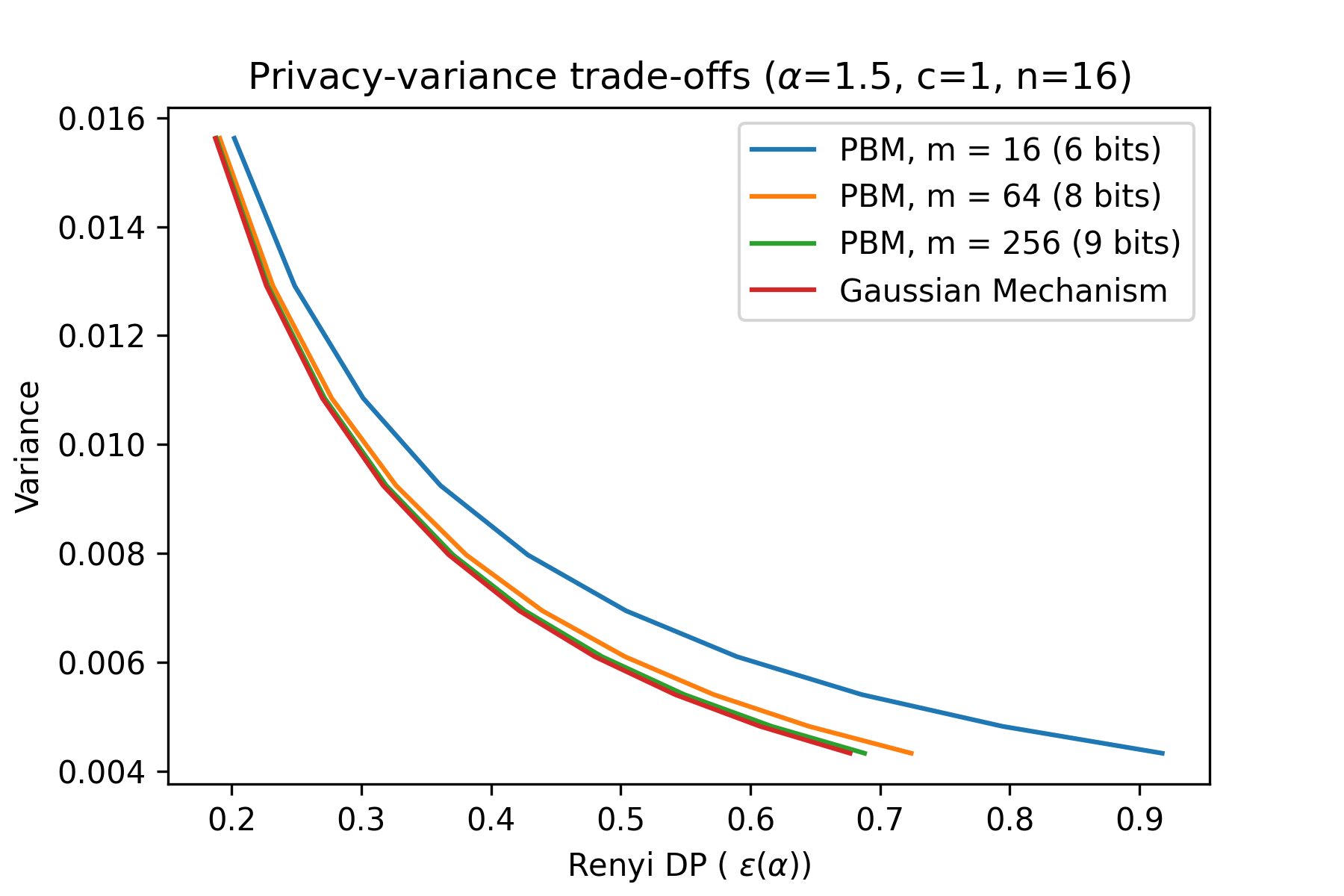} 
  	\caption{ Privacy-MSE (variance) trade-offs of PBM and the Gaussian mechanism.}%
  	\label{fig:pbm_m}
\end{figure}

In Figure~\ref{fig:pbm_m}, we numerically compute the privacy guarantee of Algorithm~\ref{alg:pbm} and compared it with the Gaussian mechanism. For the PBM, we fix the communication cost (i.e. fix $m$), vary parameter $\theta$, and compute the corresponding R\'enyi DP (i.e., $\varepsilon(\alpha)$) and MSE. We see that as $m$ increases, the privacy-MSE curve approaches to that of the Gaussian mechanism, indicating that our scheme is also optimal in its leading constant. We present another numerical results in Section~\ref{sec:numerical_appendix}, in wich we fix $\theta$ and vary $m$ to get the trade-off curves.

\section{The Scalar Poisson Binomial Mechanism}\label{sec:scalar_pbm}
In this section, we analyze the utility and privacy guarantees of the scalar version of PBM (i.e., with $d =1$). Recall that when $d=1$, each $x_i$ in the DME task (see the formulation in Section~\ref{sec:formulation}) becomes a bounded real number with $|x_i| < c$ eliminating the Kashin step. Under this special case,  Algorithm~\ref{alg:scalar_pbm} encodes each $x_i$ into a parameter of a binomial distribution by 1) first mapping $x_i$ into $\lb \frac{1}{2}-\theta, \frac{1}{2}+\theta \rb$ by  $p_i \eqDef \frac{1}{2}+\frac{\theta}{c}x_i$, and then 2) generating a binomial random variable $Z_i \sim \msf{Binom}(m, p_i)$.

Notice that from each $Z_i$, one can obtain an unbiased estimator on $x_i$ by computing 
$ \hat{x_i} = \frac{c}{\theta}\lp \frac{1}{m}Z_i - \frac{1}{2}\rp.$
Therefore, upon collecting $\sum_{i} Z_i$ from SecAgg protocol, the server can estimate $\mu$ by
$ \hat{\mu}\lp \sum_i Z_i \rp \eqDef \frac{c}{nm\theta}\lp \sum_i Z_i- \frac{mn}{2}\rp $
(recall that the server can only learn $\sum_i Z_i$ but not individual $Z_i$).

\begin{remark}
As discussed before, $m$ and $\theta$ can be chosen to achieve a desired privacy-utility-communciation trade-off. Intuitively, with larger $m$, one can reduce the variance of the estimator while weakening the privacy guarantees; similarly with smaller $\theta$, one would get a better privacy guarantee by trading off the accuracy. 
\end{remark}

\paragraph{Utility of the scalar PBM.}
As mentioned above, $\hat{\mu}$ yields an \emph{unbiased} estimate on $\mu$, and the variance can be calculated as
    $\Var\lp \hat{\mu} \rp 
    = \frac{c^2}{m^2\theta^2}\sum_i \Var\lp Z_i \rp 
     \leq \frac{c^2}{4nm\theta^2}.$

On the other hand, since $Z_i \leq m$, $\sum_i Z_i \leq nm$. Thus to avoid overflow, we will set $M = nm$, where recall that $M$ the size of the finite group SecAgg operates on. Therefore, the communication cost of Algorithm~\ref{alg:scalar_pbm} is $\log M = \log n + \log m$ bits per client.

\paragraph{Privacy of the scalar PBM.}
Next, the privacy guarantee (in an RDP form), is summarized in the following corollary.
\begin{corollary}\label{cor:rdp}
Let $m \in \mbb{N}$ and $\theta \in [0, \frac{1}{4}]$ be parameters of Algorithm~\ref{alg:scalar_pbm}. Then Algorithm~\ref{alg:scalar_pbm} satisfies $(\alpha, \varepsilon(\alpha))$-RDP for any $\alpha > 1$ and
\begin{equation}
    \varepsilon(\alpha) \geq C_0 \lp \frac{\theta^2}{(1-2\theta)^4}\rp\frac{\alpha m}{n},
\end{equation}
where $C_0 > 0$ is an universal constant.
\end{corollary}


\subsection{Analysis of the RDP}
To analyze the privacy loss of Algorithm~\ref{alg:pbm}, let $Y_i \sim \msf{Binom}(m, p_i)$ and $Y_1'\sim \msf{Binom}(m, p_1')$, where recall that $p_1,...,p_n, p_1' \in \lb \frac{1}{2}-\theta, \frac{1}{2}+\theta \rb$.  For any $\alpha > 1$, the $\varepsilon(\alpha)$ is given by
\begin{equation}\label{eq:binom_original_obj}
\max_{p_1,...,p_n, p_1'} D_\alpha\lp P_{Y_1+Y_2+...+Y_n}\mV P_{Y'_1+Y_2+...+Y_n} \rp,
\end{equation} 
with the maximum taken over $\lb \frac{1}{2}-\theta, \frac{1}{2}+\theta \rb^{n+1}$.
Our main technical contribution is the following (orderwise) tight upper bound on the  R\'enyi divergence of two Poisson binomial distributions, which then characterizes the privacy loss of our scheme.
\begin{theorem}\label{thm:rdp}
Let $\alpha > 1$ and $p_1,...,p_n, p'_1 \in [\frac{1}{2}-\theta, \frac{1}{2}+\theta]$. Let $Y_i \sim \msf{Binom}(m, p_i)$ and $Y_1'\sim \msf{Binom}(m, p_1')$. Then it holds that
\begin{align*}
    &D_\alpha\lp P_{Y_1+Y_2+...+Y_n}\mV P_{Y'_1+Y_2+...+Y_n} \rp\\
    &\vspace{3em}\leq C_0  \frac{\theta^2}{(1-2\theta)^4}\lp\min\lp4, \frac{\alpha^2}{\alpha-1}\rp \rp\frac{m}{n},
\end{align*}
where $C_0 > 0$ is an universal constant.
\end{theorem}
\begin{remark}
Although we present an asymptotic result here, using the quasi-convexity of R\'enyi divergence, one can show that the worst-case scenario is attained by the extremal points (i.e., when $p_i \in \lbp \frac{1}{2}-\theta, \frac{1}{2}+\theta \rbp$), as shown in Lemma~\ref{lemma:quasi_cvx}. This allows us to efficiently compute the privacy loss \emph{exactly}, as shown in Section~\ref{sec:numerical}.
\end{remark}

An immediate corollary of Theorem~\ref{thm:rdp} is the RDP guarantee of the proposed PBM (summarized in Corollary~\ref{cor:rdp}).

In the rest of this section, we provide a proof of Theorem~\ref{thm:rdp}. 
\paragraph{\bf{Step 0: decomposing $Y_i$}.}  To begin with, observe that since $Y_i \sim \msf{Binom}(m, p_i)$, we can decompose it into sum of $m$ independent and identical copies of $\msf{Ber}(p_i)$, i.e., $Y_i = \sum_{j=1}^m X^{(j)}_i$, where $X^{(j)}_i \diid \msf{Ber}(p_i)$ for $j \in [m]$. Therefore 
$$ \sum_{i=1}^n Y_i = \sum_{i=1}^n \sum_{j = 1}^m X^{(j)}_i = \sum_{j=1}^m \underbrace{\lp  \sum_{i=1}^n X^{(j)}_i\rp}_{\eqDef Z_j},$$
and similarly we can write $Y_1' + \sum_{i=2}^n Y_i = \sum_{j=1}^m Z_j'$, where $Z_j' = X^{'(j)}_1+\sum_{i=2}^n X^{(j)}_i$. 

Grouping the summation of $X^{(j)}_i$ according to $j \in [m]$ and applying the data processing inequality for  R\'enyi divergence, we upper bound \eqref{eq:binom_original_obj} by
\begin{align}\label{eq:ber_obj}
&\max_{p_1,...,p_n, p_1'}D_\alpha\lp P_{Y_1+Y_2+...+Y_n}\mV P_{Y'_1+Y_2+...+Y_n} \rp \nonumber\\
&= \max_{p_1,...,p_n, p_1'}D_\alpha\lp P_{Z_1+...+Z_m}\mV P_{Z'_1...+Z'_m} \rp \nonumber\\
&\leq \max_{p_1,...,p_n, p_1'}mD_\alpha\lp P_{Z_1}\mV P_{Z'_1} \rp \nonumber\\
& = \max_{p_1,...,p_n, p_1'}mD_\alpha\lp P_{X_1+X_2+...+X_n}\mV P_{X'_1+X_2+...+X_n} \rp,
\end{align}
where $X_i \sim \msf{Ber}(p_i)$ and $X'_1 \sim \msf{Ber}(p'_1)$.

\paragraph{Step 1: maximum achieved by extremal points.} 
Next, since $(P, Q) \mapsto D_\alpha\lp P \mV Q\rp$ is quasi-convex \citep[Thoerem~13]{van2014renyi}, we claim that \eqref{eq:ber_obj} is maximized at extreme points:
\begin{lemma}\label{lemma:quasi_cvx}
\eqref{eq:ber_obj} is maximized at extreme points 
i.e., when $ p_1,...,p_n, p_1' \in \lbp \frac{1}{2}-\theta, \frac{1}{2}+\theta \rbp$.
\end{lemma}

This implies \eqref{eq:ber_obj} can be upper bounded by the following binomial form:
\begin{align}\label{eq:ber_extreme}
&\max_{k \in [n-1]} D_\alpha\Big( P_{\msf{Binom}\lp 1+k, \frac{1}{2}-\theta\rp + \msf{Binom}\lp n-k-1, \frac{1}{2}+\theta\rp}\big\Vert P_{\msf{Binom}\lp k, \frac{1}{2}-\theta\rp + \msf{Binom}\lp n-k, \frac{1}{2}+\theta\rp} \Big).
\end{align}

\begin{figure}[h]
    \centering
    \includegraphics[width=0.45\linewidth]{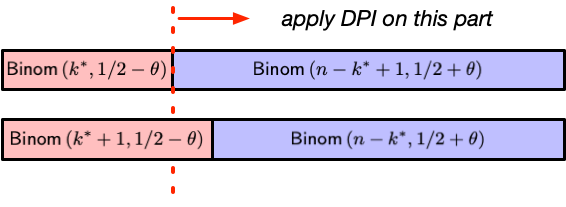}
    \caption{An illustration of applying data processing inequality, where under the scenario of $k^* \leq \frac{1}{2}$, we discard the first half of common binomial sum.}
    \label{fig:pbm_dpi}
\end{figure}

\paragraph*{\bf{Step 2: applying data processing inequality}.} 
Next, we simplify \eqref{eq:ber_extreme} by carefully applying the data processing inequality. Let $k^* \in [n-1]$ maximize \eqref{eq:ber_extreme}. If $k^* \leq \frac{n}{2}$, we apply the data processing inequality to discard the first half of common binomial random variables (see Figure~\ref{fig:pbm_dpi} for an illustration), i.e.,
$$ \msf{Binom}\lp k^*, \frac{1}{2}-\theta\rp+\msf{Binom}\lp n'-k^*, \frac{1}{2}+\theta\rp,$$ 
where $n' \eqDef \lceil \frac{n-1}{2} \rceil$. On the other hand, if $k^* \geq \frac{n}{2}$, then we apply the data processing inequality to remove the second half of common parts, i.e.,
$$ \msf{Binom}\lp n'+k^*, \frac{1}{2}-\theta\rp+\msf{Binom}\lp n-k^*-1, \frac{1}{2}+\theta\rp.$$

This leads to the following lemma:
\begin{lemma}\label{lemma:dpi}
\eqref{eq:ber_extreme} is upper bounded by the maximum of the following two quantities:
\begin{enumerate}[(a)]
    \item $D_\alpha\lp P_{\msf{Ber}\lp \frac{1}{2}-\theta\rp + \msf{Binom}\lp n', \frac{1}{2}+\theta\rp} \mV P_{\msf{Binom}\lp n'+1, \frac{1}{2}+\theta\rp} \rp$,
    \item $D_\alpha\lp P_{\msf{Binom}\lp n'+1, \frac{1}{2}-\theta\rp} \mV P_{ \msf{Ber}\lp \frac{1}{2}+\theta\rp+\msf{Binom}\lp n', \frac{1}{2}-\theta\rp} \rp$.
\end{enumerate}
\end{lemma}

\paragraph*{\bf{Step 3: bounding the  R\'enyi divergence via MGF}.} 
Finally, we upper bound each of the two terms in \eqref{eq:two_binom_bdd} separately. We start with the following simple but useful lemma, which bounds the  R\'enyi divergence of two distributions by the sub-Gaussian norm of their likelihood ratio (LR). 

\begin{lemma}\label{lemma:subgaussian_bdd}
Let $P, Q$ be two probability measures on $\mcal{X}$ and let $\frac{dP}{dQ}(x)$ be the Radon-Nikodym derivative. Let $X \sim Q$. Then for any $\alpha > 1$,
$$ D_\alpha\lp P \mV Q \rp \leq C_0 \frac{\alpha^2}{\alpha-1}\lV \frac{dP}{dQ}\lp X \rp -1\rV^2_{\psi_2},  $$
where $\lV Z \rV_{\psi_2}$ denotes the sub-Gaussian norm of $Z$ and $C_0 > 0$ is a universal constant.
\end{lemma}

To apply Lemma~\ref{lemma:subgaussian_bdd} to control (a) and (b) in Lemma~\ref{lemma:dpi}, we need to compute and bound the sub-Gaussian norms of the likelihood ratio (LRs)
of random variables in (a) and (b) of Lemma~\ref{lemma:dpi}, respectively. 

To this end, let us define 
$ R(i) \eqDef\frac{P_{\msf{Ber}\lp \frac{1}{2}-\theta\rp + \msf{Binom}\lp n', \frac{1}{2}+\theta\rp}(i)}{P_{\msf{Binom}\lp n'+1, \frac{1}{2}+\theta\rp}(i)}. $
Then the LRs corresponding to random variables of (a) and (b) in Lemma~\ref{lemma:dpi} are $R(I)$ and $1/ R(I')$ respectively, where $I \sim P_{\msf{Binom}\lp n'+1, \frac{1}{2}+\theta\rp}$ and $ I' \sim P_{ \msf{Ber}\lp \frac{1}{2}-\theta\rp+\msf{Binom}\lp n', \frac{1}{2}+\theta\rp}$. 

It turns out that $R(i)$ is a linear function of $i$, and since both $I$ and $I'$ are sum of binary random variables, one can control their sub-gaussian norms (and hence that of $R(I)$ and $1/R(I')$). We summarize the bound in the following lemma and defer the proof to Section~\ref{proof:LR_subgaussian_bdd}.

\begin{lemma}\label{lemma:LR_subgaussian_bdd}
Let $R(i)$ be defined as above and let $I \sim P_{\msf{Binom}\lp n'+1, \frac{1}{2}+\theta\rp}$ and $ I' \sim P_{ \msf{Ber}\lp \frac{1}{2}-\theta\rp+\msf{Binom}\lp n', \frac{1}{2}+\theta\rp}$. Then 
\begin{itemize}
    \item $\lV R(I) - 1\rV^2_{\psi_2} \leq C_1 \frac{\theta^2}{(1-4\theta^2)^2(n'+1)},$
    \item $\lV 1/R(I')-1 \rV^2_{\psi_2}\leq C_2 \frac{\theta^2}{(1-2\theta)^4}\frac{1}{n'+1},$
\end{itemize}
for some $C_1, C_2 > 0$.
\end{lemma}

\paragraph{Step 4: putting everything together.}
Combining Lemma~\ref{lemma:LR_subgaussian_bdd}, Lemma~\ref{lemma:subgaussian_bdd}, and Lemma~\ref{lemma:dpi}, we obtain that
\begin{align*}
    & \eqref{eq:ber_extreme}\leq C_3\frac{\theta^2}{(1-2\theta)^4}\lp \frac{\alpha^2}{\alpha-1} \rp\frac{1}{n'+1}.
\end{align*}
Together with Lemma~\ref{lemma:quasi_cvx} and \eqref{eq:ber_obj}, we conclude that 
\begin{align}\label{eq:rdp_bdd_final}
    \max_{p_1,...,p_n, p_1'}&D_\alpha\lp P_{Y_1+Y_2+...+Y_n}\mV P_{Y'_1+Y_2+...+Y_n} \rp\nonumber\\
    &\leq C_7 \lp \frac{\theta^2}{(1-2\theta)^4}\rp\lp \frac{\alpha^2}{\alpha-1} \rp\frac{m}{n'+1}\nonumber\\
    &\leq C_0 \lp \frac{\theta^2}{(1-2\theta)^4}\rp\lp \frac{\alpha^2}{\alpha-1} \rp\frac{m}{n},
\end{align}
for some $C_0 > 0$ large enough. Finally, since  R\'enyi divergence is increasing with $\alpha$, we also have for $\alpha < 2$,
\begin{align}\label{eq:rdp_bdd_final2}
    \max_{p_1,...,p_n, p_1'}D_\alpha\lp P_{Y_1+Y_2+...+Y_n}\mV P_{Y'_1+Y_2+...+Y_n} \rp
    &\leq \max_{p_1,...,p_n, p_1'}D_2\lp P_{Y_1+Y_2+...+Y_n}\mV P_{Y'_1+Y_2+...+Y_n} \rp \nonumber\\
    &\leq C_0 \lp \frac{\theta^2}{(1-2\theta)^4}\rp\frac{4m}{n}.
\end{align}
Combining \eqref{eq:rdp_bdd_final} and \eqref{eq:rdp_bdd_final2}, we establish Theorem~\ref{thm:rdp}.

\section{The Multi-dimensional PBM}\label{sec:multi_pbm}
Next, we extend the scalar PBM into the multi-dimensional setting, where $x_i \in \mbb{R}^d$ and $\lV x_i \rV_2 \leq c$. The description of multi-dimensional PBM is given in Algorithm~\ref{alg:pbm}. The key step that allows us to cast the multi-dimensional DME into the scalar one is via Kashin's representation, which transforms the $\ell_2$ geometry of the data into an $\ell_\infty$ geometry and hence enables us to decompose the problem into scalar sub-tasks.

\subsection{Kashin's representation}
We first introduce the idea of a tight frame in Kashin's representation. A tight frame is a set of vectors $\lbp u_j\rbp^D_{j=1} \in \mbb{R}^d$ that satisfy Parseval's identity, i.e.
$ \left\| x \right\|^2_2 = \sum_{j=1}^D \lan u_j, x \ran^2 \, \text{ for all } x \in \mbb{R}^d.$

A frame can be viewed as a generalization of the notion of an orthogonal basis in $\mbb{R}^d$ for $D>d$. 
To increase robustness, we wish the information to be spread evenly across different coefficients, which motivates the following  definition of  a Kashin's representation:
\begin{definition}[Kashin's representation\cite{kashin1977section}]
    For a set of vectors $\lbp u_j\rbp_{j=1}^D$, we say the expansion 
    $$ x = \sum_{j=1}^D a_ju_j, \text{ with } \max_j \lba a_j \rba \leq \frac{K}{\sqrt{D}}\lV x \rV_2 $$
    is a Kashin's representation of vector $x$ at level $K$ . 
\end{definition}


By Theorem~3.5 and Theorem~4.1 in \cite{lyubarskii2010uncertainty}, we have the following lemma:
\begin{lemma}[Uncertainty principle]\label{lemma:up_Kashin}
There exists a tight frame $U = [u_1,...,u_D]$ with (1) $D = \Theta(d)$ and (2) Kashin's level $K = O\lp 1\rp$. 
\end{lemma}

Lemma~\ref{lemma:up_Kashin} implies that for each $x_i \in \mbb{R}^d$ such that $\lV x_i \rV_2\leq 1$, one can always represent each $x_i$ with coefficients $y_i \in [-\gamma_0/\sqrt{d}, \gamma_0/\sqrt{d}]^{\gamma_1 d}$ for some  $\gamma_0, \gamma_1 > 0$ and $x_i = Uy_i$.

\subsection{Proof of Theorem~\ref{thm:main_informal}}
With Lemma~\ref{lemma:up_Kashin}, we are well-prepared to analyze the performance of Algorithm~\ref{alg:pbm}. Recall that the three main steps in the multi-dimensional PBM are:
\begin{enumerate}
    \item (Clients) compute a Kashin's representation of $x_i$ with respect to a (common) tight frame $U$ (denoted as $y_i$).
    \item (Clients) sequentially transmit each coordinate of $y_i$ via the scalar PBM.
    \item (Server) reconstructs $\mu$ by $\hat{\mu} = U\hat{\mu}_y$.
\end{enumerate}

Let us denote $y_i(j)$ as the $j$-th coordinate of $y_i$ for $j \in [\gamma_1 d]$, and $\mu_y(j)$ as the $j$-th coordinate of $\mu_y = \frac{1}{n}\sum_{i=1}^n y_i$. Due to the property of the Kashin's representation, we know that if $\lV x_i\rV_2 \leq c$, $\lV y_i \rV_\infty \leq \frac{\gamma_0 c}{\sqrt{d}}$. 

Therefore, using the scalar PBM (with parameters $\theta, m$) for coordinate $j$  and applying Theorem~\ref{thm:rdp}, it holds that
\begin{itemize}
    \item the privacy loss is $\varepsilon_{\msf{j}}(\alpha) = C_0\lp \frac{\theta^2}{(1-2\theta)^4} \rp\frac{\alpha m}{n}$;
    \item $\E\lb \hat{\mu}_y(j)\rb = \hat{\mu}_y(i)$ and 
    $ \E\lb \lp \hat{\mu}_y(j)-\mu_j \rp^2\rb \leq \frac{\gamma^2_0c^2}{4dnm\theta}; $
    \item The communication cost is $O\lp \log n + \log m\rp$ bits.
\end{itemize}

Repeating for $j=1,..,\gamma_1 d$  and accounting the overall privacy loss via the composition theorem of RDP \citep[Proposition~1]{mironov2017renyi}, the end-to-end RDP guarantee of Algorithm~\ref{alg:scalar_pbm} becomes $ \varepsilon(\alpha) = \sum_{j}\varepsilon_j(\alpha) = \gamma_1C_0\frac{d\alpha m\theta^2}{(1-2\theta)^4n}. $
Similarly, the communication cost is $\gamma_1 d\lp \log n + \log m \rp$ bits. 
Finally, we control the $\ell_2$ estimation error $\E\lb\lV \mu - \hat{\mu}\rV^2_2\rb $. Note that since $x_i = U y_i$ for all $i = 1,...,n$, we have
$ \mu_x = \frac{1}{n}\sum_i Uy_i = U\mu_y. $
Also,
\begin{align*}
    &\E\lb \lp \hat{\mu}  - \mu  \rp^2 \rb 
      = \E\lb\lV \sum\nolimits_{j} \lp \hat{\mu}_y(j)-\mu_y(j)\rp u_j \rV_2^2 \rb\\ 
     &\overset{\text{(a)}}{\leq}  \E\lb  \sum\nolimits_{j} \lp \hat{\mu}_y(j)-\mu_y(j)\rp^2 \rb
      =  \E\lb \lV \hat{\mu}_y - \mu_y \rV_2^2 \rb,
\end{align*}
where (a) is due to the Cauchy–Schwarz inequality.
Hence to bound the MSE of $\hat{\mu}$, it suffices to bound
$ \E\lb \Vert \hat{\mu}_y - \mu_y \Vert^2_2\rb \leq \gamma_1\gamma_0^2\frac{c^2}{4mn\theta} = O\lp \frac{c^2}{4mn\theta} \rp. $

\section{Empirical evaluations}

In this section, we evaluate PBM on the distributed mean estimation (DME) task. We follow the set-up of \citet{kairouz2021distributed, agarwal2021skellam}: we generate $n=1000$ client vectors with dimension $d = 250$, i.e., $x_1,...,x_n \in \mbb{R}^{250}$ . Each local vector has bounded $\ell_2$ and $\ell_\infty$ norms, i.e. $\lV x_i \rV_2 \leq 1$ and $\lV x_i \rV_\infty \leq \frac{1}{\sqrt{d}}$\footnote{We perform the experiments under an $\ell_\infty$ geometry. Under $\ell_2$ geometry, one can either apply random rotation and $\ell_\infty$ clipping, or compute the Kashin's representation, as discussed in Section~\ref{sec:multi_pbm}.}. Our goal is to
demonstrate that the utility of PBM matches that of the continuous Gaussian mechanism
under the same privacy guarantees when given sufficient communication budget.

\begin{figure}[ht]
    \centering
    \includegraphics[width=0.55\linewidth]{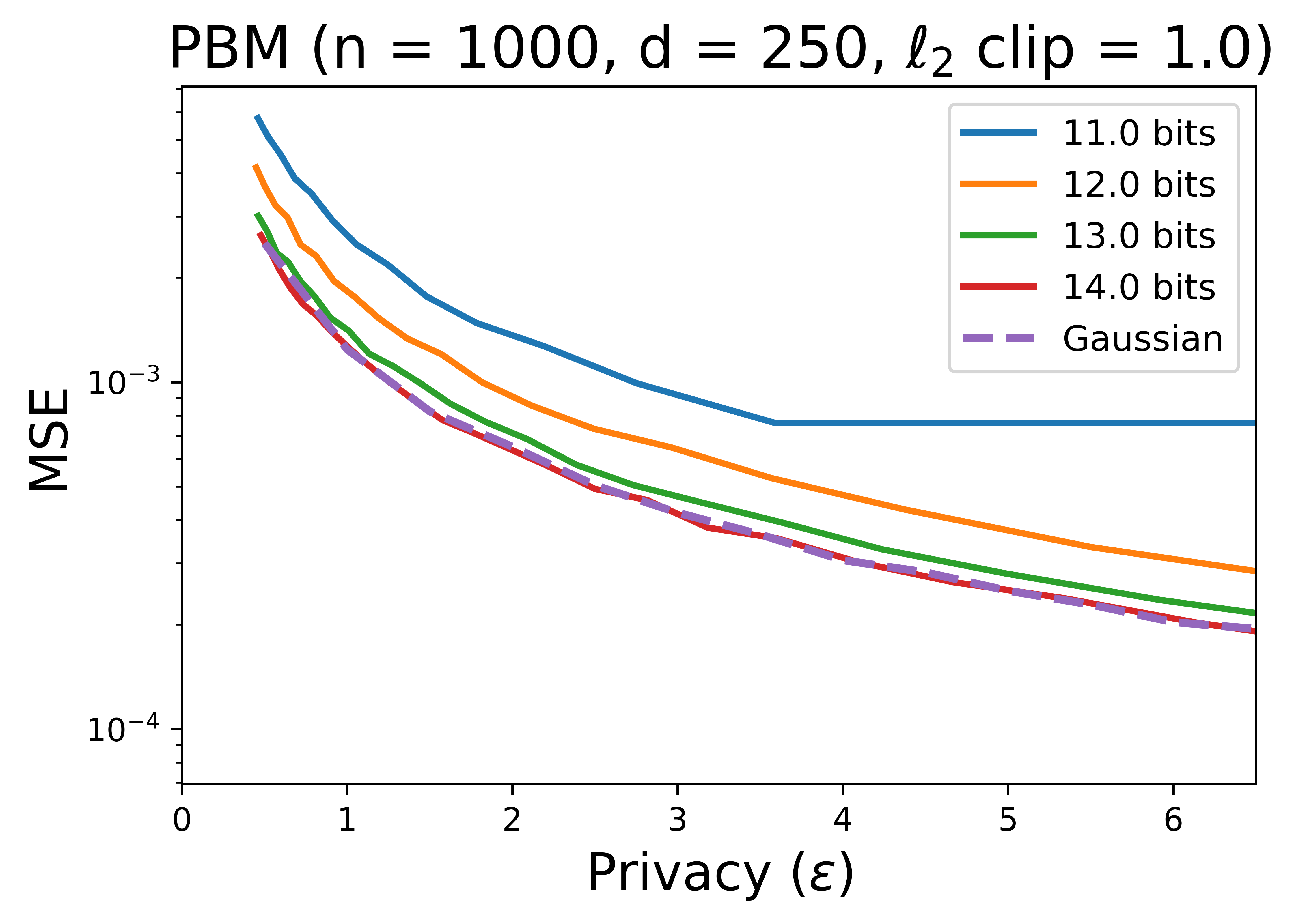}
    \caption{A comparison of PBM with the continuous Gaussian mechanism. We set $m = \{2, 4, 6, 16\}$, and the corresponding communication costs (i.e., the logarithmic of the field size that SecAgg operates on) are $B = \{ 11, 12, 13, 14\}$.}
    \label{fig:dme_no_clipping}
\end{figure}

In Figure~\ref{fig:dme_no_clipping}, we apply PBM with different $m$'s (which dictate the communication cost) under a given privacy requirement $\varepsilon$. Note that once $m$ is determined, the field that SecAgg operates on will be $\mbb{Z}_{2^{\lceil \log_2(n\cdot m)\rceil}}$ and hence the communication cost becomes $\lceil \log_2(n\cdot m)\rceil$ bits. 

\begin{figure}[ht]
    \centering
    \includegraphics[width=0.65\linewidth]{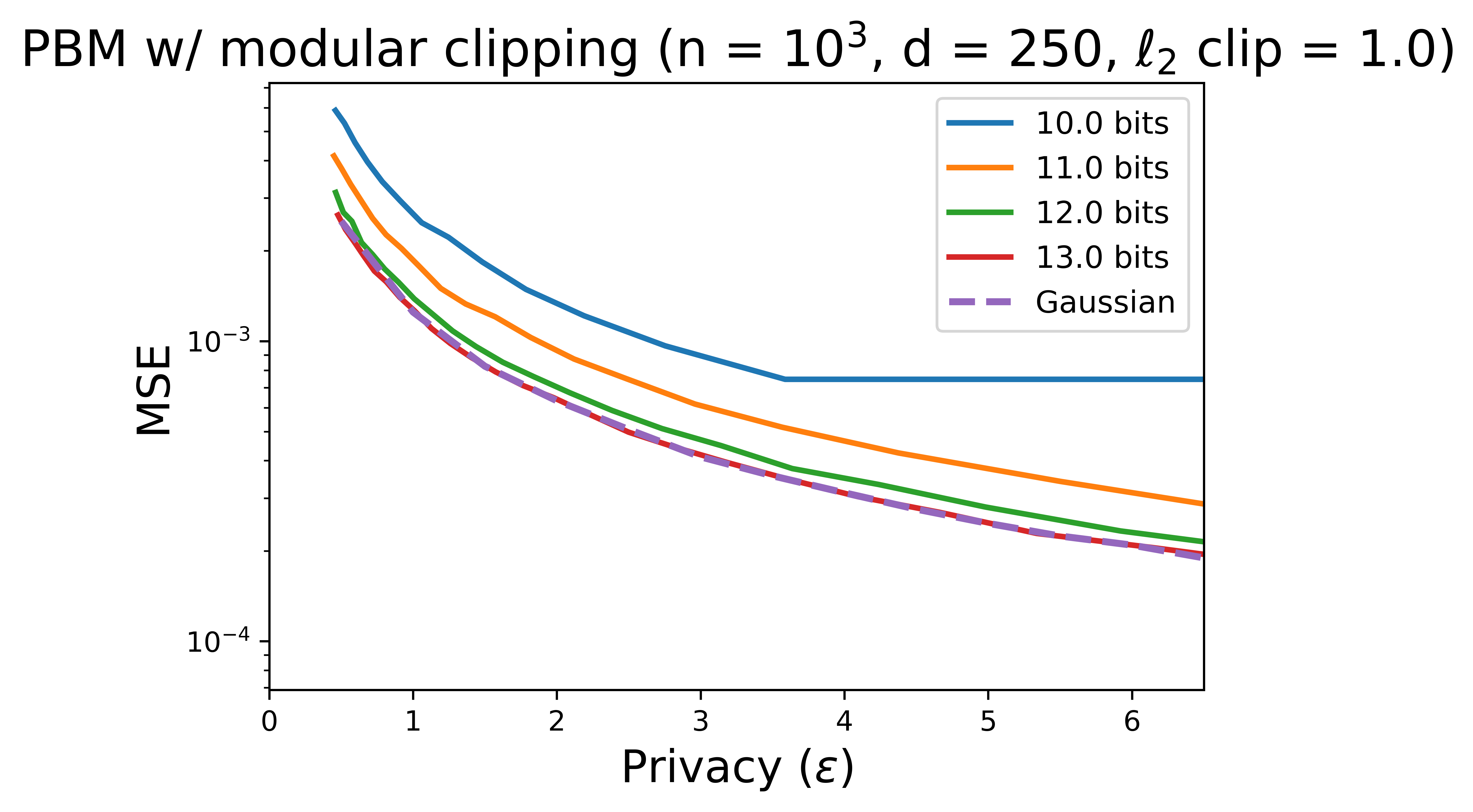}
    \caption{PBM with modular clipping, where $m = \{2, 4, 6, 16\}$. We set the modular clippiing parameter $c = \sqrt{30}$ and report the MSEs and the corresponding communication cost.}
    \label{fig:dme_clipping}
\end{figure}

\paragraph{Reducing communication via modular clipping.} Since the sum of encoded messages (i.e., $\sum Z_i$ in Algorithm~\ref{alg:scalar_pbm}) follows a Poisson-binomial distribution with $p_i \in \lb \frac{1}{2}-\theta, \frac{1}{2}+\theta \rb$, with high-probability,  $\sum Z_i \in \lb \frac{nm(1-\theta)}{2} - c\sqrt{\frac{nm}{4}}, \frac{nm(1+\theta)}{2} + c\sqrt{\frac{nm}{4}}\rb$ (where $c$ is a parameter used to control the failure probability). Therefore one can (modularly) clip local message in that range to further reduce and hence communication cost the field size to $nm\theta + c\sqrt{nm}$ (this, however, will incur \emph{bias} to the final estimator). In Figure~\ref{fig:dme_clipping}, we perform PBM with modular clipping and set $c = \sqrt{30}$. Since $c$ is large enough, we observe almost no impact to the errors while saving the (per-parameter) communication by $1$ bit.

\section{Application to private SGD}
In this section, we apply PBM to distributed SGD. In each round, the server samples $n$ out of $N$ clients randomly, each (sampled) client computes a local gradient from its data, and the server aggregates the mean of the local gradients via PBM. Since PBM ensures distributed DP, we call the resulting scheme DDP-SGD.

We summarize DDP-SGD in Algorithm~\ref{alg:ddp_sgd}, in which we use $\msf{PBM}_{\texttt{enc}}$ to denote the clients' procedure in Algorithm~\ref{alg:pbm} (which includes computing Kashin's representation, sequentially applying the scalar PBM, and performing SecAgg) and $\msf{PBM}_{\texttt{dec}}$ to denote the server's procedure.

\begin{algorithm}[tbh]
   \caption{Distributed DP-SGD}
   \label{alg:ddp_sgd}
\begin{algorithmic}
   \STATE {\bfseries Input:} Clients data set $d_1,...,d_N \in \mcal{D}$, PBM parameters $(\theta, m)$, loss function $\ell(d, w)$.
   \STATE {\bfseries Goal:} Compute $w_T \approx \arg\min_{w} \sum_{i=1}^N\ell\lp d_i, w \rp$.
    \STATE Server generates an initial model weights $w_0 \in \mbb{W}$
       \FOR{iteration $t = 1,...,T$}
   \STATE Server samples a subset of $n$ clients $\mcal{C}_t\subset [N]$ and broadcasts $w_{t-1}$ to them.
   \FOR{each client $i \in \mcal{C}_t$}
    \STATE Computes $g^t_i = \msf{Clip}_{\ell_2, c}\lp \nabla \ell\lp d_i, w_{t-1} \rp\rp$
    \STATE Computes $Z^t_i = \msf{PBM}_\texttt{enc}\lp g^t_i \rp$
    \STATE Send $Z^t_i$ to the server via SecAgg
    \ENDFOR
    \STATE (Server) decodes $\hat{\mu}^t_g = \msf{PBM}_{\texttt{dec}}\lp \sum_i Z_i^t \rp$.
    \STATE (Server) updates the model by $w_t = w_{t-1}+\gamma\hat{\mu}^t_g$.
   \ENDFOR
   
\STATE {\bfseries Return:} $w_T$
\end{algorithmic}
\end{algorithm}

\paragraph{Analysis of convergence rates.}
Next, we analyze the convergence rate of Algorithm~\ref{alg:ddp_sgd}. Due to the unbiased nature of PBM, one can easily control the convergence rate of DDP-SGD by the variance of PBM through the following lemma (which originates from \cite{ghadimi2013stochastic} but we use a version adapted from \cite{agarwal2018cpsgd}):
\begin{lemma}[Corollary~1 in \cite{agarwal2018cpsgd}]\label{lemma:sgd_convergence}\label{lemma:sgd_convergence_rate}
    Assume $F(w) \eqDef \frac{1}{N}\sum_{i=1}^N \ell( w; d_i)$, where $\ell(\cdot, d)$ is an $L$-smooth and $c$-Lipschitz function for all $d \in \mcal{D}$. Let $ w_0$ satisfies $F\lp  w_0\rp - F\lp  w^* \rp \leq D_F$. Let $\hat{\mu}^t_g$ be the noisy model updates at round $t$ 
    and let the learning rate $\gamma \eqDef \min\lbp L^{-1}, \sqrt{2D_F}\lp \sigma \sqrt{LT} \rp^{-1} \rbp$. Then after $T$ rounds,
    $$ \E_{t\sim \msf{unif}(T)}\lb \lV \nabla F( w_t) \rV^2_2 \rb \leq \frac{2D_FL}{T}+\frac{2\sqrt{2}\sigma\sqrt{LD_F}}{\sqrt{T}} + cB,$$
    
    \begin{align*}
        \text{where } \sigma^2 = 2\Big( &\max_{1\leq t \leq T} \E\lb \lV \mu_g^t -\nabla F( w_t) \rV^2_2\rb\\
        &\hspace{5em}+\max_{1\leq t\leq T}\E_{Q}\lb \lV \mu_g^t - \tilde{\mu}_g^t \rV^2_2 \rb\Big), 
    \end{align*}
    and $B = \max_{1\leq t \leq T} \lV \E_Q\lb \mu_g^t - \hat{\mu}_g^t \rb \rV_2. $
\end{lemma}
Note that in each round, (1) $\mu^t_g$ (the true mean of (the sampled) clients' gradients) is an unbiased estimator of $\nabla F(w_t)$ (because clients are sampled uniformly at random), and (2) $\hat{\mu}^t_g$ is an unbiased estimator of $\mu^t_g$ since PBM is unbiased\footnote{Notice that the clipping step in Algorithm~\ref{alg:ddp_sgd} does not increase bias since by the Lipschitz condition, $\lV\nabla \ell\rV_2 \leq c$.}. This implies $B=0$ and $\sigma^2 = \max_t \Var\lp \mu_g^t \rp +\Var\lp \hat{\mu}_g^t \mv \mu_g^t \rp$, where the first term is bounded by $c^2$, and applying Theorem~\ref{thm:main_informal}, we can bound the second $\Var\lp\hat{\mu}_g^t \mv \mu_g^t \rp$ by $ \frac{c^2}{4nm\theta^2}$. Thus we arrive at the following conclusion:
\begin{corollary}[Convergence of DDP-SGD]\label{cor:ddp_sgd_convergence}
Under the same assumptions of Lemma~\ref{lemma:sgd_convergence_rate}, after $\tau \sim \msf{uniform}(T)$ iterations, the output of Algorithm~\ref{alg:ddp_sgd} satisfies
\begin{equation*}
    \E_{\tau}\lb \lV \nabla F(w_\tau) \rV^2_2\rb \leq \frac{LD_F}{T}+\frac{\sqrt{8c^2LD_F}}{\sqrt{T}}\sqrt{1+ \frac{1}{4nm\theta^2}}.
\end{equation*}
\end{corollary}
\begin{remark}
Note that due to the convergence guarantees of Lemma~\ref{lemma:sgd_convergence_rate}, in Corollary~\ref{cor:ddp_sgd_convergence} we apply Algorithm~\ref{alg:ddp_sgd} with a random stopping time.
\end{remark}

\paragraph{Accounting for total privacy loss.}
To account for the total privacy loss, we first note that the per-round RDP guarantee is amplified by the sub-sampling of the clients with sampling rate $\kappa = \frac{n}{N}$. To quantify the tight amplification rate, one can apply \citep[Theorem~9]{wang2019subsampled} and obtain a non-asymptotic upper bound on $\varepsilon_\msf{sampled}(\alpha)$. For instance, \citep[Theorem~9]{wang2019subsampled} shows that when $\alpha$ is not too large (e.g., when $\alpha \leq 2$), 
$ \varepsilon_\msf{sampled}(\alpha) = O\lp \kappa^2\varepsilon(\alpha) \rp $.
Applying Theorem~\ref{thm:main_informal} and plugging $\varepsilon(\alpha) = \frac{dm\theta^2\alpha}{n}$ in this bound, we have
$ \varepsilon_\msf{sampled}(\alpha) = O\lp \frac{ndm\theta^2\alpha}{N^2} \rp. $
To account for the privacy loss over all the $T$ iterations, we apply the composition theorem for RDP \citep[Proposition~1]{mironov2017renyi}, concluding that Algorithm~\ref{alg:ddp_sgd} satisfies $ \lp\alpha, \varepsilon_{\msf{final}}\rp $ with $\varepsilon_{\msf{final}}(\alpha) = O\lp \frac{ndm\theta^2\alpha T}{N^2} \rp$ when $\alpha\leq 2$ .

For high-privacy (where $nm\theta^2 \ll 1$) and small $\alpha$ regimes, we obtain a convergence rate-privacy trade-off:
$$ \E_{\tau}\lb \lV \nabla F(w_\tau) \rV^2_2\rb  \approx O\lp\sqrt{\frac{c^2d\alpha}{N^2\varepsilon_\msf{final}\lp\alpha\rp}}\rp.$$

This achieves the optimal rate of reported in \citep[Table~1]{bassily2014private} (though admittedly, our results only hold for high privacy regime and for $\alpha < 2$). Finally, we remark to obtain non-asymptotic trade-offs for full $\alpha > 1$ regimes, one has to resort to \citep[Theorem~9]{wang2019subsampled}.

\section{Conclusion}
In this paper, we present the Poisson Binomial mechanism, a discrete (R\'enyi) DP mechanism that can be combined with SecAgg for distributed mean estimation or federated learning/analytics. Unlike previous schemes, our mechanism is not based on additive noise, so in addition to achieving the optimal privacy-accuracy trade-off, it offers two extra advantages: (1) it results in an unbiased estimator, and (2) the communication cost with SecAgg decreases with the privacy budget $\varepsilon$. Leveraging the unbiasedness property, we propose a distributed DP-SGD algorithm and analyze its convergence rate. Several important open problems include deriving a lower bound on the communication cost for DME with DP and SecAgg, and evaluating our scheme on real-world FL datasets.

\section*{Acknowledgments}
This work was supported in part by NSF Award \# NeTS-1817205 and CCF-2213223, a Google Faculty research award, and a National Semiconductor Corporation Stanford Graduate Fellowship. The authors would like to thank Thomas Steinke for an inspiring discussion in the early stage of this work, and Graham Cormode for pointing out the  connection of PBM to \citet{cheu2019distributed} and insightful comments on an earlier draft of this paper.

\newpage
\bibliography{references}
\bibliographystyle{icml2022}

\newpage
\appendix
\onecolumn

\section{Conversion of RDP to approximate DP}\label{sec:conversion}
The following conversion lemma from \cite{asoodeh2020better, canonne2020discrete, bun2016concentrated} relates RDP to $\lp\varepsilon_{\msf{DP}}(\delta), \delta\rp$-DP.

\begin{lemma}\label{lemma:rdp_to_approx_dp}
    If $M$ satisfies $\lp \alpha, \varepsilon(\alpha) \rp$-RDP for all $\alpha > 1$, then, for any $\delta > 0$, $M$ satisfies $\lp \varepsilon_{\msf{DP}}(\delta), \delta \rp$-DP, where
$$ \varepsilon_{\msf{DP}}(\delta) = \inf_{\alpha > 1}\varepsilon(\alpha)+\frac{\log\lp 1/\alpha\delta \rp}{\alpha-1}+\log(1-1/\alpha). $$
\end{lemma}

To compare the privacy-utility trade-off of our mechanism with previous works, it will be useful sometimes to convert the stronger RDP guarantees to an approximate DP guarantee. The following lemma, which is a simple application of Lemma~\ref{lemma:rdp_to_approx_dp}, provides a simple form for the resulting approximate DP guarantees.
\begin{lemma}\label{lemma:rdp_to_approx_dp_2}
If $M$ satisfies $\lp \alpha, \varepsilon(\alpha) \rp$-RDP for all $\alpha > 1$, then, for any $\delta > 0$, $M$ satisfies $\lp \varepsilon_{\msf{DP}}(\delta), \delta \rp$-DP, where
$$ \varepsilon_{\msf{DP}}(\delta) = \Theta\lp \sqrt{\lp\sup_{\alpha} \frac{\varepsilon(\alpha)}{\alpha}\rp\log(1/\delta)} \rp. $$
\end{lemma}

\section{Additional plots}\label{sec:numerical_appendix}

\begin{figure}[ht]
  	\centering
  	\includegraphics[width=0.6\linewidth]{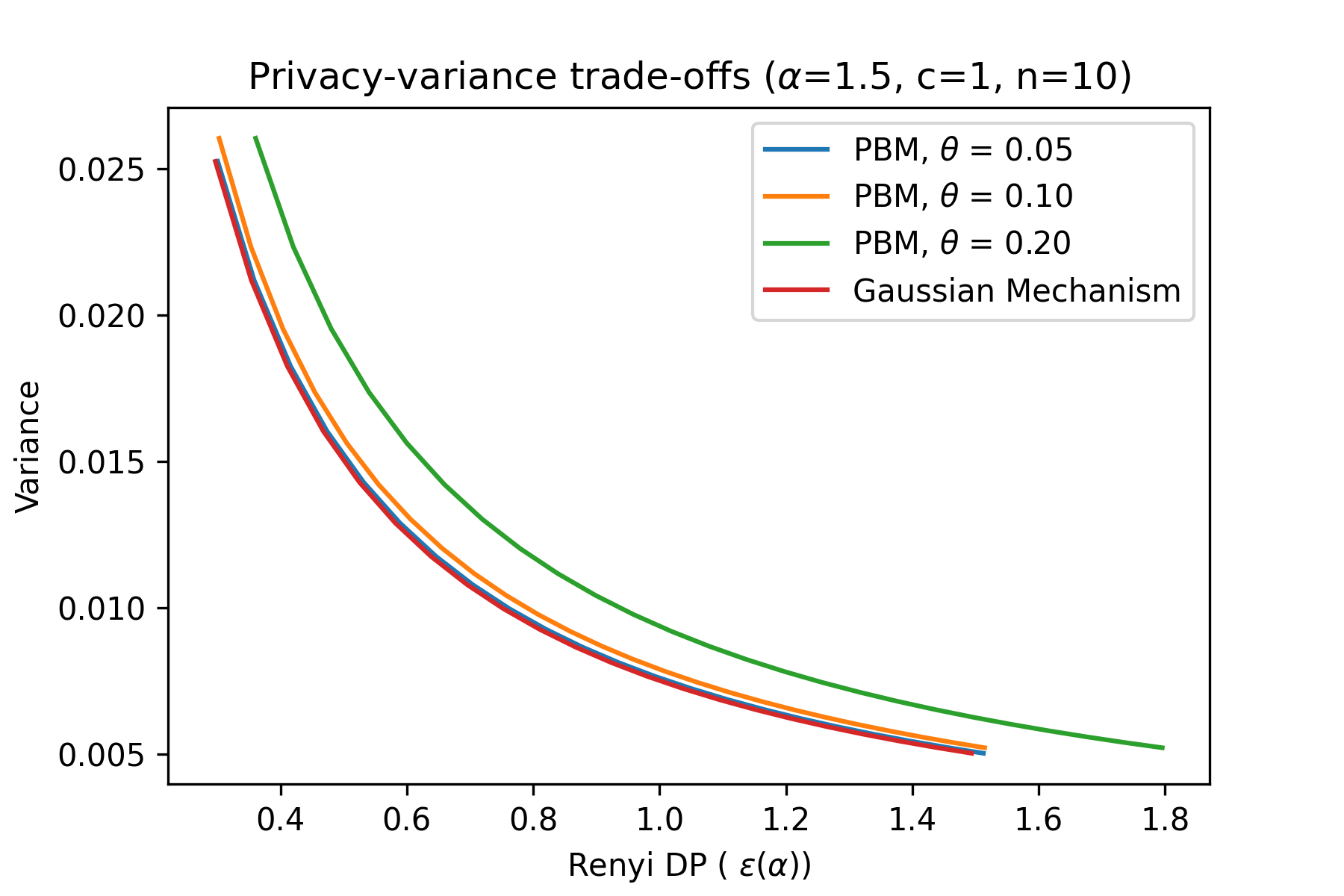} 
  	\caption{ Privacy-MSE (variance) trade-offs of PBM and the Gaussian mechanism.}%
  	\label{fig:pbm_theta}
\end{figure}

\paragraph{Numerical evaluation of the scalar PBM.} In figure~\ref{fig:pbm_theta}, we compute the privacy guarantee of Algorithm~\ref{alg:pbm} and compared it with the Gaussian mechanism. For the PBM, we fix $\theta$, vary parameter m, and compute the corresponding R\'enyi DP (i.e., $\varepsilon(\alpha)$) and MSE. We see that as $\theta \ra 0$, the privacy-MSE curve converges to that of the Gaussian mechanism fast.

\section{Additional Proofs}
\subsection{Proof of Lemma~\ref{lemma:rdp_to_approx_dp_2}}
\begin{align*}
    \varepsilon_{\msf{DP}}(\delta) 
    & \leq \varepsilon(\alpha)+\frac{\log\lp 1/\alpha\delta \rp}{\alpha-1}+\log(1-1/\alpha)\\
    & \leq \varepsilon(\alpha) +\frac{\log(1/\delta)}{\alpha-1} \\
    &  = \frac{\varepsilon(\alpha)}{\alpha} + \frac{\varepsilon(\alpha)}{\alpha}(\alpha-1)+\frac{\log(1/\delta)}{\alpha-1} .
\end{align*}

Next, take $\alpha^* = 1+\sqrt{\sup_{\alpha}\frac{\alpha}{\varepsilon(\alpha)}}$, we get
$$ \varepsilon_{\msf{DP}}(\delta) \leq \sup_{\alpha}\frac{\varepsilon(\alpha)}{\alpha} + 2\sqrt{\lp\sup_{\alpha} \frac{\varepsilon(\alpha)}{\alpha}\rp\log(1/\delta)} = \Theta\lp \sqrt{\lp\sup_{\alpha} \frac{\varepsilon(\alpha)}{\alpha}\rp\log(1/\delta)} \rp, $$
if $\frac{\varepsilon(\alpha)}{\alpha} = O(1)$.
This suggests that $(\alpha, \varepsilon(\alpha))$-RDP implies $(\varepsilon_{\msf{DP}}(\delta), \delta)$-DP with $\varepsilon_{\msf{DP}}(\delta) = \Theta_\delta\lp \sqrt{\lp\sup_{\alpha} \frac{\varepsilon(\alpha)}{\alpha}\rp} \rp$.

\subsection{Proof of Lemma~\ref{lemma:quasi_cvx}}
To see this, observe that $P_{X_1+X_2+...+X_n} = P_{X_1+...+X_{n-1}}\circ P_{X_n}$, where $\circ$ denotes the convolution operator. Since the convolution operator is linear, we have
$$ P_{X_1+...+X_{n-1}}\circ P_{X_n} = \lambda \lp P_{X_1+...+X_{n-1}} \circ \msf{Ber}\lp \frac{1}{2}-\theta\rp\rp +(1-\lambda)\lp P_{X_1+...+X_{n-1}} \circ \msf{Ber}\lp\frac{1}{2}+\theta\rp\rp, $$
where $\lambda > 0$ is such that $p_n = \lambda\lp \frac{1}{2}-\theta \rp + (1-\lambda)\lp \frac{1}{2}+\theta \rp$. By the quasi-convexty of $D_\alpha(\cdot \Vert \cdot)$, it holds that $D_\alpha\lp P_{X_1+X_2+...+X_n}\mV P_{X'_1+X_2+...+X_n} \rp$ is upper bounded by
$$  
\max\lp D_\alpha\lp P_{X_1+X_2+...+\msf{Ber}\lp\frac{1}{2}-\theta \rp}\mV P_{X'_1+X_2+...+\msf{Ber}\lp\frac{1}{2}-\theta \rp} \rp, D_\alpha\lp P_{X_1+X_2+...+\msf{Ber}\lp\frac{1}{2}+\theta \rp}\mV P_{X'_1+X_2+...+\msf{Ber}\lp\frac{1}{2}+\theta \rp} \rp \rp.$$

By repetitively applying the quasi-convexity and the extrema argument for $n-1$ times, we arrive at
\begin{align*}
    &D_\alpha\lp P_{X_1+X_2+...+X_n}\mV P_{X'_1+X_2+...+X_n} \rp
    \leq \max_{k \in [n-1]}D_\alpha\lp P_{X_1+N_k}\mV P_{X'_1+N_k} \rp,
\end{align*}
where $N_k \sim \msf{Binom}(k, \frac{1}{2}-\theta)+\msf{Binom}(n-k-1, \frac{1}{2}+\theta)$.
    
In the last step, by making use of the \emph{joint} quasi-convexity of R\'enyi divergence (i.e., $(P, Q) \mapsto D_\alpha(P \Vert Q)$ is quasi-convex), it suffices to show that
\begin{align*}
    &\lp P_{X_1+N_k}, P_{X'_1+N_k} \rp = \lambda^*_1\lp P_{\msf{Ber}\lp \frac{1}{2}-\theta \rp+N_k}, P_{\msf{Ber}\lp \frac{1}{2}-\theta \rp+N_k} \rp\\
    &+\lambda^*_2 \lp P_{\msf{Ber}\lp \frac{1}{2}-\theta \rp+N_k}, P_{\msf{Ber}\lp \frac{1}{2}+\theta \rp+N_k} \rp
    + \lambda^*_3\Big( P_{\msf{Ber}\lp \frac{1}{2}+\theta \rp+N_k}, \\ 
    &P_{\msf{Ber}\lp \frac{1}{2}-\theta \rp+N_k} \Big)
    +\lambda^*_4\lp P_{\msf{Ber}\lp \frac{1}{2}+\theta \rp+N_k}, P_{\msf{Ber}\lp \frac{1}{2}+\theta \rp+N_k} \rp
\end{align*}
for some $\lambda^*_i \in [0, 1]$ and $\sum_{i=1}^4 \lambda^*_i = 1$. To this end, since $X_i \sim \msf{Ber}(p_i)$ and $X'_i \sim \msf{Ber}(p'_i)$ for some $p_i, p'_i \in \lb \frac{1}{2}-\theta, \frac{1}{2}+\theta \rb$, we must have 
$$ P_{X_1 + N_k} = \lambda P_{\msf{Ber}\lp \frac{1}{2}-\theta \rp+N_k}+(1-\lambda) P_{\msf{Ber}\lp \frac{1}{2}+\theta \rp+N_k},  \text{ and}$$
$$ P_{X'_1 + N_k} = \lambda' P_{\msf{Ber}\lp \frac{1}{2}-\theta \rp+N_k}+(1-\lambda') P_{\msf{Ber}\lp \frac{1}{2}+\theta \rp+N_k}  $$
for some $\lambda, \lambda' \in [0, 1]$.
Therefore, by setting $\lambda_1^* = \lambda\lambda'$, $\lambda_2^* = \lambda(1-\lambda')$, $\lambda_3^* = (1-\lambda)\lambda'$, and $\lambda_4^* = (1-\lambda)(1-\lambda')$, we arrive at the desired result.

\subsection{Proof of Lemma~\ref{lemma:dpi}}
\begin{align*}
    & P_{\msf{Binom}\lp 1+k^*, \frac{1}{2}-\theta\rp + \msf{Binom}\lp n-k^*-1, \frac{1}{2}+\theta\rp} = P_{\msf{Ber}\lp \frac{1}{2}-\theta\rp + \msf{Binom}\lp n-k^*-1, \frac{1}{2}+\theta\rp} \circ  P_{\msf{Binom}\lp k^*, \frac{1}{2}-\theta\rp}\\
    & P_{\msf{Binom}\lp k^*, \frac{1}{2}-\theta\rp + \msf{Binom}\lp n-k^*, \frac{1}{2}+\theta\rp} = P_{ \msf{Binom}\lp n-k^*, \frac{1}{2}+\theta\rp} \circ  P_{\msf{Binom}\lp k^*, \frac{1}{2}-\theta\rp},
\end{align*}
and by a data processing inequality, we have 
\begin{align*}
    &D_\alpha\lp P_{\msf{Binom}\lp 1+k^*, \frac{1}{2}-\theta\rp + \msf{Binom}\lp n-k^*-1, \frac{1}{2}+\theta\rp} \mV P_{\msf{Binom}\lp k^*, \frac{1}{2}-\theta\rp + \msf{Binom}\lp n-k^*, \frac{1}{2}+\theta\rp} \rp\\
    &\leq 
    D_\alpha\lp P_{\msf{Ber}\lp \frac{1}{2}-\theta\rp + \msf{Binom}\lp n-k^*-1, \frac{1}{2}+\theta\rp} \mV P_{ \msf{Binom}\lp n-k^*, \frac{1}{2}+\theta\rp} \rp\\
    &\leq 
    D_\alpha\lp P_{\msf{Ber}\lp \frac{1}{2}-\theta\rp + \msf{Binom}\lp n', \frac{1}{2}+\theta\rp} \mV P_{ \msf{Ber}\lp \frac{1}{2}+\theta\rp+\msf{Binom}\lp n', \frac{1}{2}+\theta\rp} \rp,
\end{align*}
where in the last inequality we use a data processing inequality and define $n' = \lceil \frac{n-1}{2}\rceil$. Similarly, when $k^* > \frac{n}{2}$, we decompose 
\begin{align*}
    & P_{\msf{Binom}\lp 1+k^*, \frac{1}{2}-\theta\rp + \msf{Binom}\lp n-k^*-1, \frac{1}{2}+\theta\rp} = P_{ \msf{Binom}\lp 1+k^*, \frac{1}{2}-\theta\rp} \circ  P_{\msf{Binom}\lp n-k^*-1, \frac{1}{2}+\theta\rp}\\
    & P_{\msf{Binom}\lp k^*, \frac{1}{2}-\theta\rp + \msf{Binom}\lp n-k^*, \frac{1}{2}+\theta\rp} = P_{ \msf{Ber}\lp \frac{1}{2}+\theta\rp +\msf{Binom}\lp k^*, \frac{1}{2}-\theta\rp} \circ  P_{\msf{Binom}\lp n-k^*-1, \frac{1}{2}+\theta\rp}.
\end{align*}
By a data processing inequality, we obtain
\begin{align*}
    &D_\alpha\lp P_{\msf{Binom}\lp 1+k^*, \frac{1}{2}-\theta\rp + \msf{Binom}\lp n-k^*-1, \frac{1}{2}+\theta\rp} \mV P_{\msf{Binom}\lp k^*, \frac{1}{2}-\theta\rp + \msf{Binom}\lp n-k^*, \frac{1}{2}+\theta\rp} \rp\\
    &\leq 
    D_\alpha\lp P_{\msf{Ber}\lp \frac{1}{2}-\theta\rp + \msf{Binom}\lp n', \frac{1}{2}-\theta\rp} \mV P_{ \msf{Ber}\lp \frac{1}{2}+\theta\rp+\msf{Binom}\lp n', \frac{1}{2}-\theta\rp} \rp.
\end{align*}
Therefore we conclude that \eqref{eq:ber_extreme} can be upper bounded by
\begin{align}\label{eq:two_binom_bdd}
    &\max_{k \in [n-1]}D_\alpha\lp P_{\msf{Binom}\lp 1+k, \frac{1}{2}-\theta\rp + \msf{Binom}\lp n-k-1, \frac{1}{2}+\theta\rp} \mV P_{\msf{Binom}\lp k, \frac{1}{2}-\theta\rp + \msf{Binom}\lp n-k, \frac{1}{2}+\theta\rp} \rp\nonumber\\
    &\leq \max\Big(
    \underbrace{D_\alpha\lp P_{\msf{Ber}\lp \frac{1}{2}-\theta\rp + \msf{Binom}\lp n', \frac{1}{2}+\theta\rp} \mV P_{ \msf{Ber}\lp \frac{1}{2}+\theta\rp+\msf{Binom}\lp n', \frac{1}{2}+\theta\rp} \rp}_{\text{(a)}}\Big), \nonumber\\
    & \hspace{6em}\underbrace{D_\alpha\lp P_{\msf{Ber}\lp \frac{1}{2}-\theta\rp + \msf{Binom}\lp n', \frac{1}{2}-\theta\rp} \mV P_{ \msf{Ber}\lp \frac{1}{2}+\theta\rp+\msf{Binom}\lp n', \frac{1}{2}-\theta\rp} \rp}_{\text{(b)}}.
\end{align}

\subsection{Proof of Lemma~\ref{lemma:subgaussian_bdd}}
Using the inequality $t \leq e^{t-1}$, we have
\begin{align*}
    D_\alpha\lp P \mV Q \rp 
    & = \frac{1}{\alpha-1}\log\lp\E_Q\lb \lp\frac{dP}{dQ}\lp X \rp\rp^\alpha \rb\rp\\
    & \leq \frac{1}{\alpha-1}\log\lp\E_Q\lb e^{\alpha\lp\frac{dP}{dQ}\lp X \rp-1\rp} \rb\rp \\
    & \leq \frac{1}{\alpha-1}\log\lp\E_Q\lb e^{\alpha\lp\frac{dP}{dQ}\lp X \rp-1\rp} \rb\rp \\
    & \overset{\text{(a)}}{\leq} \frac{1}{\alpha-1}\log\lp e^{C_0\alpha^2 \lV \frac{dP}{dQ}\lp X \rp -1\rV^2_{\psi_2}} \rp \\
    & = C_0 \frac{\alpha^2}{\alpha-1}\lV \frac{dP}{dQ}\lp X \rp -1\rV^2_{\psi_2},
\end{align*}
where (a) holds for any sub-gaussian random variable (see, for instance, \citet[Proposition~2.5.2]{vershynin2018high}, which states that for a zero-mean random variable $Z$ with finite sub-gaussian norm, $\E\lb e^{\alpha Z} \rb \leq e^{C_0 \alpha^2 \lV Z\rV^2_{\psi_2}}$). 

\subsection{Proof of Lemma~\ref{lemma:LR_subgaussian_bdd}}\label{proof:LR_subgaussian_bdd}
\paragraph*{Bounding (a)} For notational simplicity, we denote the LR of term (a) in \eqref{eq:two_binom_bdd} as $R(i)$ for $i \in [n'+1]$ and bound it as follows.
\begin{align*}
    R(i) &\eqDef\frac{P_{\msf{Ber}\lp \frac{1}{2}-\theta\rp + \msf{Binom}\lp n', \frac{1}{2}+\theta\rp}(i)}{P_{\msf{Binom}\lp n'+1, \frac{1}{2}+\theta\rp}(i)}\\
    &= \frac{{n'\choose i}\lp \frac{1}{2}+\theta\rp^{i+1}\lp \frac{1}{2}-\theta\rp^{n'-i}}{{n'+1\choose i}\lp \frac{1}{2}+\theta\rp^i\lp \frac{1}{2}-\theta\rp^{n'+1-i}}
   +\frac{{n'\choose i-1}\lp \frac{1}{2}+\theta\rp^{i}\lp \frac{1}{2}-\theta\rp^{n'-i+1}}{{n'+1\choose i}\lp \frac{1}{2}+\theta\rp^i\lp \frac{1}{2}-\theta\rp^{n'+1-i}}\\
    & = \lp \frac{n'+1-i}{n'+1} \rp\lp\frac{1-2\theta}{1+2\theta}\rp + \lp\frac{i}{n'+1}\rp\lp \frac{1+2\theta}{1-2\theta} \rp \\
    & = \lp\frac{1-2\theta}{1+2\theta}\rp +\frac{i}{n'+1}\lp \lp\frac{1+2\theta}{1-2\theta}\rp-\lp\frac{1-2\theta}{1+2\theta}\rp\rp\\
    & = \lp\frac{1-2\theta}{1+2\theta}\rp +\frac{i}{n'+1}\lp \frac{8\theta}{1-4\theta^2}\rp.
\end{align*}
When $I \sim P_{\msf{Binom}\lp n'+1, \frac{1}{2}+\theta\rp}$, $I-\E[I]$ has a sub-gaussian norm $\lV I-\E[I] \rV^2_{\psi_2} \leq \sigma^2_0 (n'+1)$ for some universal constant $\sigma_0$ (since $I$ is sum of $n'+1$ independent binary random variables). Also notice that $\E[R(I)] = 1$, so $R(I)-1$, which is a linear function of $I$, can be written as 
$$ R(I) - 1 = \frac{8\theta}{(n'+1)(1-4\theta^2)}\lp I - \E[I] \rp. $$
Therefore, $R(I)-1$ has a sub-gaussian norm bounded by 
$$ \lV R(I) - 1\rV^2_{\psi_2} \leq C_1 \frac{\theta^2}{(1-4\theta^2)^2(n'+1)}.$$
By Lemma~\ref{lemma:subgaussian_bdd}, we conclude that term (a) in \eqref{eq:two_binom_bdd} can be controlled by
\begin{align}\label{eq:two_binomial_term_a_bdd}
    &D_\alpha\lp P_{\msf{Ber}\lp \frac{1}{2}-\theta\rp + \msf{Binom}\lp n', \frac{1}{2}+\theta\rp} \mV P_{\msf{Binom}\lp n'+1, \frac{1}{2}+\theta\rp} \rp \nonumber\\
    &\leq C_2\frac{\theta^2}{(1-4\theta^2)^2}\frac{\alpha^2}{\alpha-1}\frac{1}{n'+1},
\end{align}
for some constant $C_2 > 0$.

\paragraph*{Bounding (b)} Similarly, let us denote the LR of term (b) in \eqref{eq:two_binom_bdd} as $R'(i) = \frac{1}{R(i)}$ for $i \in [n'+1]$. Let $$ I \sim P_{ \msf{Ber}\lp \frac{1}{2}-\theta\rp+\msf{Binom}\lp n', \frac{1}{2}+\theta\rp}$$ 
and we control the sub-gaussian norm of $R'(I)-1$ as follows:
\begin{align}\label{eq:R'-1}
    R'(I)-1 
    = \frac{\E[R(I)]-R(I)}{R(I)} + \frac{1-\E[R(I)]}{R(I)}.
\end{align}
For the first term in the right hand side of \eqref{eq:R'-1}, observe that
\begin{align}
    \frac{\E[R(I)]-R(I)}{R(I)} &\leq \frac{\lba\E[R(I)]-R(I)\rba}{\lba R(I)\rba}\nonumber\\
    &\leq \lp\frac{1+2\theta}{1-2\theta}\rp\lba\E[R(I)]-R(I)\rba
\end{align}
where the last inequality holds since $R(I) \geq \lp\frac{1+2\theta}{1-2\theta}\rp$ almost surely. Therefore
\begin{align}\label{eq:R'_first_term_bdd}
    &\lV \frac{\E[R(I)]-R(I)}{R(I)} \rV^2_{\psi_2}\nonumber\\ 
    & \leq  \lp\frac{1+2\theta}{1-2\theta}\rp^2 \lV R(I) - \E[R(I)] \rV^2_{\psi_2}\nonumber\\
    & \leq C_3 \lp\frac{1+2\theta}{1-2\theta}\rp^2\frac{\theta^2}{(1-4\theta^2)^2}\frac{\alpha^2}{\alpha-1}\frac{1}{n'+1}\nonumber\\
    & = C_3 \frac{\theta^2}{(1-2\theta)^4}\frac{1}{n'+1},
\end{align}
where the second inequality is due to the fact that $I-\E[I]$ is sum of $n'+1$ zero-mean bounded random variables.

Next, the second term in \eqref{eq:R'-1} can be controlled by
\begin{align}\label{eq:R'_second_term_bdd}
    &\lba \frac{1-\E[R(I)]}{R(I)}\rba 
     \leq \lp\frac{1+2\theta}{1-2\theta}\rp \lba R(I) - 1 \rba\nonumber\\
    & = \lp\frac{1+2\theta}{1-2\theta}\rp\lba\lp\frac{1-2\theta}{1+2\theta}\rp +\frac{\E[I]}{n'+1}\lp \frac{8\theta}{1-4\theta^2}\rp - 1\rba\nonumber\\
    & = \lp\frac{1+2\theta}{1-2\theta}\rp\lp \frac{8\theta^2}{\lp1-4\theta^2\rp(n'+1)}\rp\nonumber\\
    & =  \frac{8\theta^2}{\lp1-2\theta\rp^2(n'+1)},
\end{align}
where the second equality holds since 
$$\E[I] = n'\lp 1+\frac{\theta}{2}\rp +\lp 1-\frac{\theta}{2}\rp. $$

Combining \eqref{eq:R'_first_term_bdd} and \eqref{eq:R'_second_term_bdd}, we obtain an upper bound on \eqref{eq:R'-1}:
\begin{align*}
    &\lV R'(I)-1 \rV^2_{\psi_2}\\
    &\leq 2\lV \frac{\E[R(I)]-R(I)}{R(I)} \rV^2_{\psi_2} +2\lV \frac{8\theta^2}{\lp1-2\theta\rp^2(n'+1)} \rV^2_{\psi_2} \\
    &\leq 2C_3\frac{\theta^2}{(1-2\theta)^4}\frac{1}{n'+1} + C_4\lp \frac{8\theta^2}{\lp1-2\theta\rp^2(n'+1)}\rp^2 \\
    & \leq C_5 \frac{\theta^2}{(1-2\theta)^4}\frac{1}{n'+1},
\end{align*}
for some $C_5 > 0$. Also notice that $ \E[R'(I) - 1] = 0 $ when $ I \sim P_{ \msf{Ber}\lp \frac{1}{2}-\theta\rp+\msf{Binom}\lp n', \frac{1}{2}+\theta\rp}$. Therefore, applying Lemma~\ref{lemma:subgaussian_bdd}, we conclude that term (b) in \eqref{eq:two_binom_bdd} can be controlled by
\begin{align}\label{eq:two_binomial_term_b_bdd}
    &D_\alpha\lp P_{\msf{Binom}\lp n'+1, \frac{1}{2}+\theta\rp} \mV P_{ \msf{Ber}\lp \frac{1}{2}-\theta\rp+\msf{Binom}\lp n', \frac{1}{2}+\theta\rp} \rp\nonumber\\ 
    &\leq C_6\frac{\theta^2}{(1-2\theta)^4}\frac{\alpha^2}{\alpha-1}\frac{1}{n'+1},
\end{align}
for some $C_6 > 0$.

\end{document}